\documentclass[10pt,journal,compsoc]{IEEEtran}
\usepackage{amsmath} 
\usepackage{amssymb}                         
\usepackage[ruled,norelsize]{algorithm2e}    
\usepackage{bm}                              
\usepackage{booktabs} 
\usepackage[caption=false]{subfig}    
\usepackage[pdftex]{graphicx}

\ifCLASSOPTIONcompsoc
  \usepackage[nocompress]{cite}
\else
  \usepackage{cite}
\fi
\ifCLASSINFOpdf
\else
\fi
\begin{document}
%
\title{MetaRadar: Indoor Localization by Reconfigurable Metamaterials}
%
%
%
%

\author{Haobo~Zhang,~\IEEEmembership{Student Member,~IEEE,}
        Jingzhi~Hu,~\IEEEmembership{Student Member,~IEEE,}
        Hongliang~Zhang,~\IEEEmembership{Member,~IEEE,}
        Boya~Di,~\IEEEmembership{Member,~IEEE,}
        Kaigui~Bian,~\IEEEmembership{Senior Member,~IEEE,}
        Zhu~Han,~\IEEEmembership{Fellow,~IEEE,}
        and~Lingyang~Song,~\IEEEmembership{Fellow,~IEEE}
\IEEEcompsocitemizethanks{\IEEEcompsocthanksitem H. Zhang, J. Hu, and L. Song are with Department of Electronics, Peking University, Beijing, China (e-mail: \{haobo.zhang,jingzhi.hu,lingyang.song\}@pku.edu.cn).

\IEEEcompsocthanksitem H. Zhang is with Department of Electronics, Peking University, Beijing, China, and also with Department of Electrical and Computer Engineering, University of Houston, TX, USA (e-mail: hongliang.zhang92@gmail.com).

\IEEEcompsocthanksitem B. Di is with Department of Electronics, Peking University, Beijing, China, and also with Department of Computing, Imperial College London, London, UK (e-mail: diboya92@gmail.com).

\IEEEcompsocthanksitem K. Bian is with Department of Computer Science, Peking University, Beijing, China (e-mail: bkg@pku.edu.cn).

\IEEEcompsocthanksitem Z. Han is with the Department of Electrical and Computer Engineering, University of Houston, Houston, TX 77004, USA, and also with the Department of Computer Science and Engineering, Kyung Hee University, Seoul 17104, South Korea (e-mail: zhan2@uh.edu).
}
}

\IEEEtitleabstractindextext{%
\begin{abstract}
  Indoor localization has drawn much attention owing to its potential for supporting location based services. Among various indoor localization techniques, the received signal strength (RSS) based technique is widely researched. However, in conventional RSS based systems where the radio environment is unconfigurable, adjacent locations may have similar RSS values, which limits the localization precision. In this paper, we present MetaRadar, which explores reconfigurable radio reflection with a surface/plane made of metamaterial units for multi-user localization. By changing the reflectivity of metamaterial, MetaRadar modifies the radio channels at different locations, and improves localization accuracy by making RSS values at adjacent locations have significant differences. However, in MetaRadar, it is challenging to build radio maps for all the radio environments generated by metamaterial units and select suitable maps from all the possible maps to realize a high accuracy localization. To tackle this challenge, we propose a compressive construction technique which can predict all the possible radio maps, and propose a configuration optimization algorithm to select favorable metamaterial reflectivities and the corresponding radio maps. The experimental results show a significant improvement from a decimeter-level localization error in the traditional RSS-based systems to a centimeter-level one in MetaRadar.
\end{abstract}

\begin{IEEEkeywords}
  Indoor localization, reconfigurable intelligent surface, received signal strength.
\end{IEEEkeywords}}

\maketitle

\IEEEdisplaynontitleabstractindextext

%
\IEEEpeerreviewmaketitle

\IEEEraisesectionheading{\section{Introduction}\label{sec:introduction}}

%
%
%
%
\IEEEPARstart{L}{ocation-based} services, such as navigation and mobile recommender system, are considered as an indispensable part of many emerging applications in Mobile Internet~\cite{zafari2019survey}. In outdoor environments, location information with acceptable accuracy can be provided by Global Positioning System (GPS), while the localization in indoor scenarios is still challenging mainly due to the uncontrollable multi-path effects~\cite{mainetti2014survey}. 

To enable indoor location based services, various localization techniques are proposed, among which the technique based on received signal strength (RSS) in wireless local area network (WLAN) has gained much attention~\cite{yassin2016recent}. Leveraging the widespread WLAN infrastructure in indoor environments, the RSS-based technique can provide location information for user devices equipped with existing Wi-Fi modules, thus avoiding the cost of installing any extra localization hardware. Besides, comparing with channel state information (CSI) based technique, RSS information can be easily obtained without the need of using some advanced WiFi network interface cards~\cite{ibrahim2018cnn}.

However, the performance of the RSS-based localization depends on the colleted RSS values. Specifically, a RSS-based system usually contains two phases: the \emph{offline} and the \emph{online} phases. In the first phase, the system collects a specific RSS value for each sampling location, and these values are all stored into a database, i.e., the radio map. Then, in the online phase, the system estimates the user's location by comparing the RSS value measured by the user and the stored values in the radio map~\cite{yang2012locating}. In the uncontrollable radio environment, the radio map is passively measured and cannot be customized, and the existence of adjacent locations whose RSS values in the radio map are similar to each other inevitably degrades the performance of the localization system.

Recently, metametarials have been used as a potential tool to actively configure the radio environment in the wireless communication systems~\cite{di2019smart}. A surface/plane of materials, a.k.a. metasurface, contains numerous metamaterial units with tunable reflectivities. By changing the reflectivities of metamaterial units independently, the metasurface is capable of changing the reflected radio frequency (RF) signals in a desired way, and therefore the surrounding radio environment can be customized~\cite{zhang2019reconfigurable}. This paves a new way to actively alter the radio maps and reduce the similarity of the RSS values corresponding to adjacent locations, which improves the localization accuracy.

In this paper, we propose MetaRadar, a metasurface assisted system for multi-user localization. With metamaterial, MetaRadar can generate favourable radio environment to achieve fine-grained localization. However, the integration of metasurface will significantly affect the RSS-based system and bring challenges to both the offline and the online phase.

\begin{itemize}
    \item In the first phase, MetaRadar needs to generate radio maps for all the possible radio environments created by metasurface in order to leverage the reconfiguration ability of metamaterial. Since the number of possible radio environments is large, it is challenging to build radio maps for all of them.
    \item In the second phase, MetaRadar has to select favorable radio maps from the vast number of available radio maps and combine the information collected under different radio maps to provide high precision location results, which will complicate the localization process.
\end{itemize}

To tackle these challenges, we carefully design the two phases of MetaRadar: the radio map preparation phase and the fine-grained localization phase. For the first phase, we propose a compressive technique which eliminates the all measurements containing unnecessary information. By using the received signals recorded in several critical radio environments, we can construct all the potential radio maps. Then, the MetaRadar uses an iterative approach for multi-user localization in the second phase. We propose a configuration optimization algorithm to decide the most suitable radio map and the corresponding metasurface reflectivity in each iteration. The RSS values measured by multiple users in this iteration, the corresponding radio map, and the information in previous iterations are all utilized to iteratively improve the accuracy of the estimated locations.

MetaRadar is implemented using a metasurface made of electrically tunable metamaterial units, and commercial universal software radio peripheral (USRP) devices as the transceiver and the receiver, in an indoor setting. The system performance is evaluated under different scenarios with different number of users. The experimental results show that the metasurface can largely improve the performance of the RSS based technique. For single users and multiple users in some scenarios, the localization accuracy can be up to centimeters. Specifically, for a single user with the distance of $1$m to the metasurface, MetaRadar can provide the location information with average localization error of $1$cm within $2$s.

The rest of the paper is organized as follows. In Section~\ref{s_rw}, we review the related work about existing indoor localization and metasurface techniques. 
In Section~\ref{s_so}, we describe the system architecture including the radio map preparation and the fine-grained localization phases. Details of the two phases are discussed in Section~\ref{s_rmcp} and Section~\ref{s_ilp}, respectively. We present the system implementation in Section~\ref{s_i} and show the evaluation results in Section~\ref{s_e}. Finally, we discuss the extra challenges in Section~\ref{s_d} and conclude the work in Section~\ref{s_c}.

\section{Related Work}
\label{s_rw}

\subsection{Indoor Localization Techniques}

Recent years have witnessed much interest in indoor localization systems. According to the enabling technologies, these systems can be categorized into several types: Wi-Fi~\cite{bahl2000radar}, radio frequency identification (RFID)~\cite{yang2012efficient, ma2017minding}, visible light~\cite{li2018rainbowlight, zhang2017pulsar}, and other technologies based systems~\cite{chuo2017rf, lu2018simultaneous}. Compared with other systems, Wi-Fi based system can locate every Wi-Fi compatible device without installing
extra hardware, thus becoming one of the most widespread indoor localization approaches~\cite{farid2013recent}.

Various techniques are adopted in Wi-Fi based systems such as RSS, CSI, angle-of-arrival (AoA), and time-of-arrival (ToA) techniques~\cite{liu2015a}. Among these techniques, the RSS technique is widely used because of the simplicity of measuring RSS and the minimum hardware requirements. Different localization systems have been designed by exploiting the RSS information. For example, the Radar system~\cite{bahl2000radar} uses the deterministic method to estimate the user location, where the information of the nearest neighborhood is utilized to infer the location. The authors in~\cite{castro2001probabilistic} store information about the RSS distributions from the access points and use the Bayesian Network approach to estimate the user location. In~\cite{youssef2005horus}, a joint clustering and probabilistic determination technique is proposed to tackle the noisy wireless channel and manage the computation cost. 
Besides, the authors in~\cite{yang2012locating, rai2012zee} explore methods to reduce the cost for radio map construction by integrating movement information from inertial sensors.

In the aforementioned works, the radio environment is passively adopted, and the RSS values at different locations cannot be reconfigured, which limits the localization accuracy. In comparison, the metasurface-aided system can actively customize the radio environment by changing the reflectivities of metamaterial units~\cite{di2019smart}. Utilizing the customized radio environment, we can obtain a radio map with favorable RSS values at different locations, which helps increase the localization accuracy. 
Besides, metasurface can also benefit the localization accuracy in multi-user systems, since it provides the flexibility of generating various radio maps which potentially suit for multi-user coexisting scenes.

\subsection{Metasurface in Wireless Networks}
\label{ss_metasurface}

The reflectivity of metamaterial units can be independently tuned, and therefore the metasurface is feasible to control the reflected signals and the surrounding radio environment. Hence, the metasurface is becoming increasingly popular in many wireless network designs~\cite{di2019hybrid, li2020programmable}. The authors in~\cite{liaskos2018realizing} use metasurface to alleviate the multi-path effects in the communication systems to provide high-quality wireless communications. In~\cite{huang2019reconfigurable}, the metasurface is utilized to support the downlink communication from the base station to multiple users. The practical beamforming ability of a $6$ square-meter metasurface is evaluated in~\cite{arun2019rfocus}. The metasurface is fabricated on printed circuit boards which is suitable for RF band. The experimental results show distinct improvement of signal strength and channel capacity. The aforementioned work shows the feasibility of using metasurfaces for the improving the performance of wireless networks.

\section{System Overview}
\label{s_so}

\begin{figure}[!t]
  \centering
  \subfloat[]{
    \includegraphics[width=1.6in]{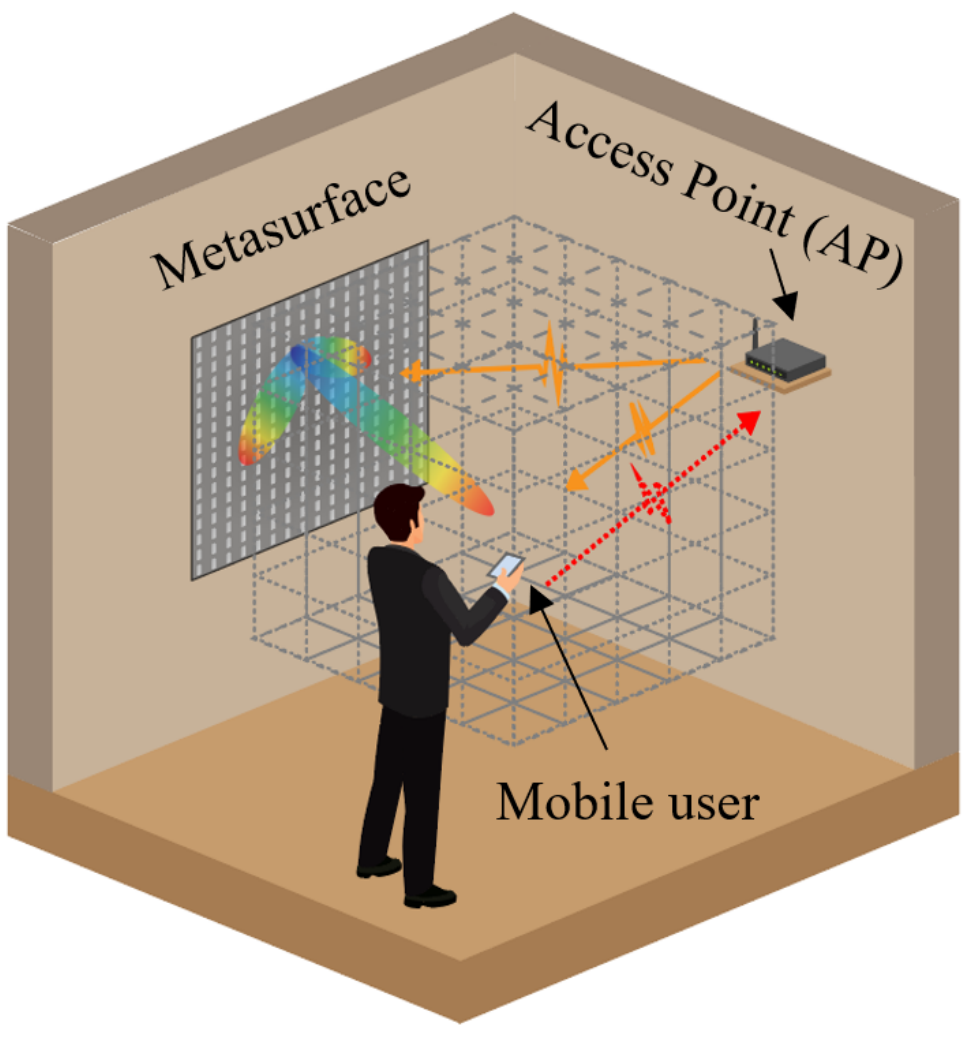}}
  \hspace{0.1in}
  \subfloat[]{
    \includegraphics[width=1.6in]{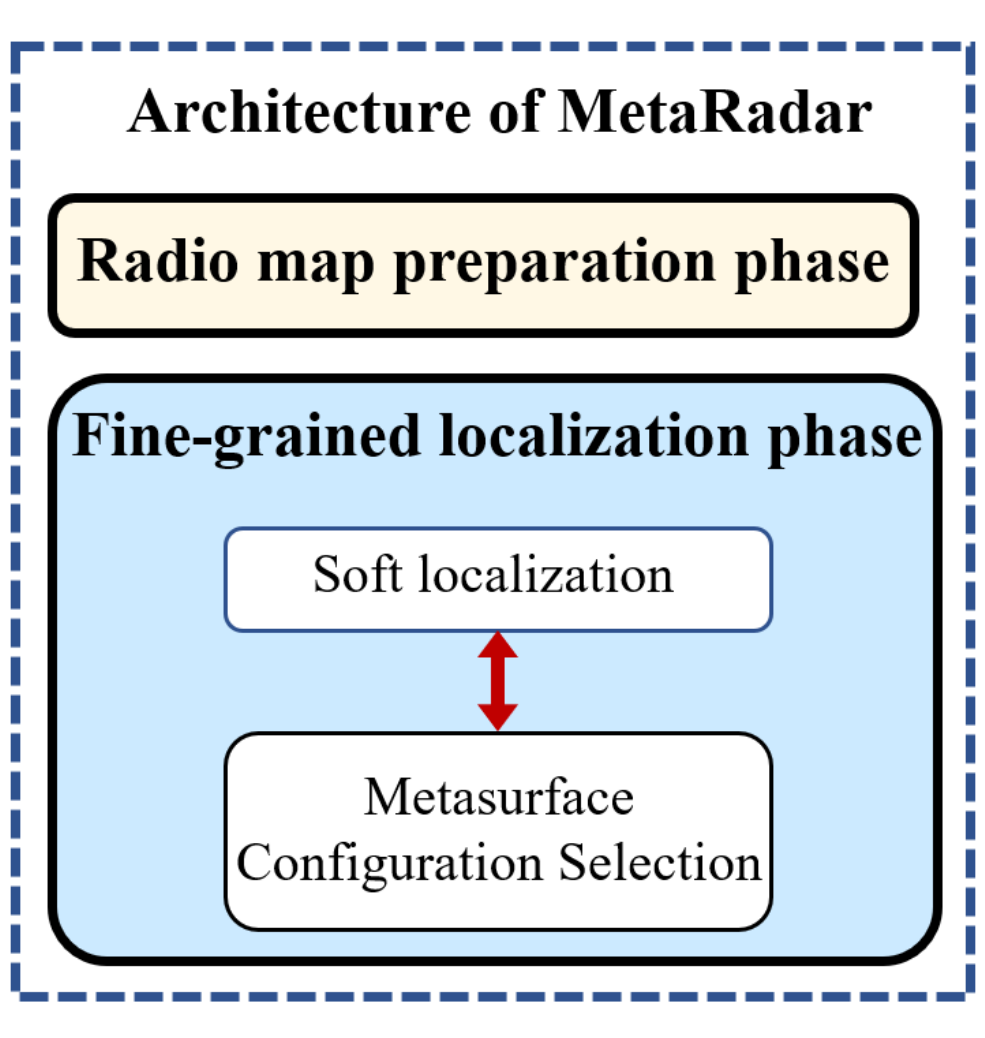}}
    
  \caption{Illustration on (a)~hardware and (b) software architectures of MetaRadar.}
  \label{f_architecture}
\end{figure}

The hardware and software architectures of the MetaRadar are presented in Figs.~\ref{f_architecture}~(a) and~(b). 
As illustrated in Fig.~\ref{f_architecture}~(a), MetaRadar is composed of a metasurface and an AP in hardware and a localization process in software, which are combined to provide location information for mobile users in indoor scenarios.
Specifically, the indoor space in MetaRadar, where the mobile users with hand-held devices to locate are, is referred to as the \emph{space of interest}~(SOI).
Without loss of generality, we assume SOI to be a cubic region which is discretized into $3$D blocks.
As shown in Fig.~\ref{f_architecture}~(b), the localization process consists of two phases: the radio map preparation phase and the fine-grained localization phase.

\textbf{Radio map preparation phase:} 
To locate multiple users, the $3$D radio map for the SOI need to be obtained in this phase.
Besides, to improve the localization capability, MetaRadar needs to select configurations of the metasurface to reconfigure the radio environment and provide favorable radio maps.
Therefore, in prior to the localization phase, MetaRadar needs to first collect RSS values in SOI given each configuration.

However, as the number of available configurations can be very large, it is costly to measure the RSS values in the SOI for all the configurations.
To address this issue, we model the received signals in the reconfigurable radio environment of MetaRadar, and then propose a compressive construction technique for MetaRadar to obtain the radio maps given each configuration, which will be discussed in Section~\ref{s_rmcp}.

\textbf{Fine-grained localization phase:}
As shown in Fig.~\ref{f_architecture}~(a), the user requiring location information sends a localization request with the current measured RSS value from her/his mobile device to the AP, which initiates the fine-grained localization.
As shown in Fig.~\ref{f_architecture}~(b), the fine-grained localization phase is composed of a soft localization and a metasurface configuration selection processes, which work iteratively.
The soft localization process takes the measured RSS value and calculates the probability for the user to be at each location.
Then, the metasurface configuration selection process optimizes the configuration of the metasurface in order to obtain a better radio map to locate the user based on the location probabilities.
The user will measure the RSS of localization signals given the newly optimized configuration of metasurface and send the RSS to the AP. The location probabilities can be utilized to estimate the locations of users, and its accuracy will be improved iteratively. The localization phase will terminate when a satisfied localization accuracy is obtained. In our experiment, each iteration takes $100$ms, and results with acceptable accuracy can be obtained within several seconds.
More details will be introduced in Section~\ref{s_ilp}. 

\section{Radio Map Preparation Phase}
\label{s_rmcp}

In this section, we first introduce the process of building the metasurface which consists of an array of metamaterial units, and then present an experimental example to show how the metasurface customizes the RSS values at a location. After that, we model the RSS which incorporates the influence of the metasurface reflection. Finally, we propose a compressive construction technique for preparing the radio map, which can predict all the possible RSS values using the measured received signal data for some critical configurations.

\subsection{Building an Array of Metamaterial Units}

\begin{figure}[!t]
  \centering
  \includegraphics[width=3.3in]{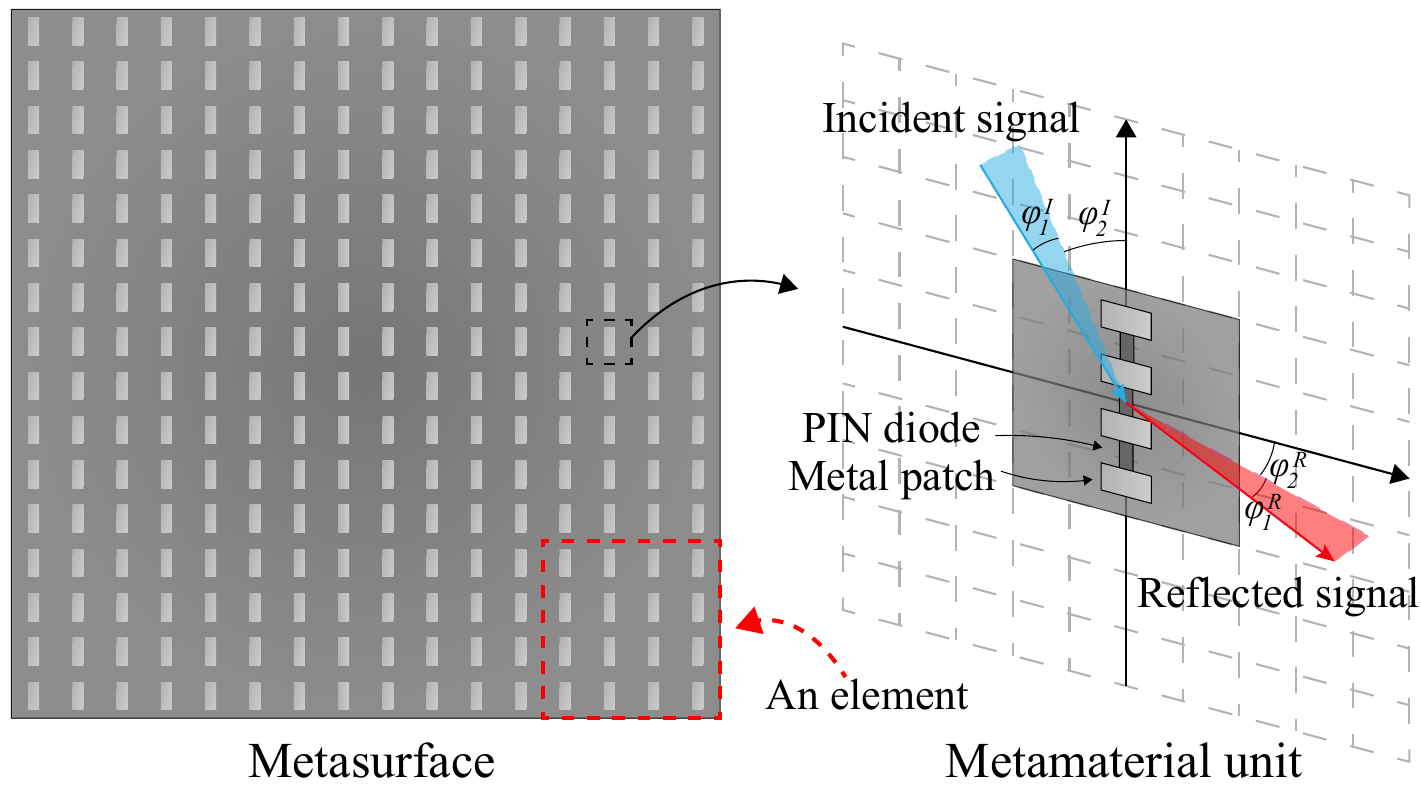}
  \caption{Reflection characteristics of an metamaterial unit.}
  \label{f_meta-atom}
\end{figure}

As illustrated in Fig.~\ref{f_meta-atom}, the metasurface is composed of an array of metamaterial units organized on a planar surface~\cite{Li2018Metasurfaces}.
Each metamaterial unit consists of several subwavelength-scale metal patches which are printed on the dielectric substrate and connected by positive-intrinsic-negative~(PIN) diodes. By applying different bias voltages, each PIN diode can be tuned into two states, i.e., \emph{ON} and \emph{OFF} states. The ON and OFF states of the PIN diodes of a metamaterial unit determine the \emph{state} of the metamaterial unit~\cite{elmossallamy2020reconfigurable}.

The metamaterial unit is able to manipulate the phase and amplitude of the reflected signals. To characterize this capability, the concept of \emph{reflectivity} is introduced, which is defined as the ratio of the reflected signals to the incident signals. Let $r$ denote the reflectivity of a metamaterial unit, which is a complex number. The amplitude of $r$, i.e., $|r|$, denotes the ratio of the reflected signal amplitude to the incident signal amplitude, and the angle of $r$, i.e., $\angle r$, is the corresponding phase shift from the incident signal to the reflected signal. The reflectivity of a metamaterial unit is determined by its state, incident angle $\bm{\varphi}^I = (\varphi^I_1, \varphi^I_2)$ of the incident signals, and reflection angle $\bm{\varphi}^R = (\varphi^R_1, \varphi^R_2)$ of the reflected signals, as shown in Fig.~\ref{f_meta-atom}. By configuring the metamaterial units into different states, the reflectivities of metamaterial units can be changed, and thus the radio environment can be modified. 

Besides, in most cases, the reflectivity is also related to the frequency of the incident signals, and only for a narrow bandwidth the reflectivities under different states are desirable~\cite{Li2018Metasurfaces}. Therefore, in this paper, we only use a single frequency sine-wave signal for localization.
We denote the frequency of the signal as $f_c$, and thus the reflectivities of the metamaterial units in different states are defined on $f_c$.

To reduce the mutual coupling among metamaterial units as well as the regulation burden~\cite{Li2019Machine}, we group adjacent metamaterial units together, which is referred to as an element. The element is the minimal unit of the metasurface which can be controlled independently. All the metamaterial units of one element are in the same state, which is referred to as the state of the element. The set of elements in the metasurface is denoted by $\mathcal{M}$. We define the \emph{configuration} of the metasurface as the states of all the elements.

\subsection{Changing the RSS Value at a Location}

\begin{figure}[!t]
  \centering
  \includegraphics[width=3.4in]{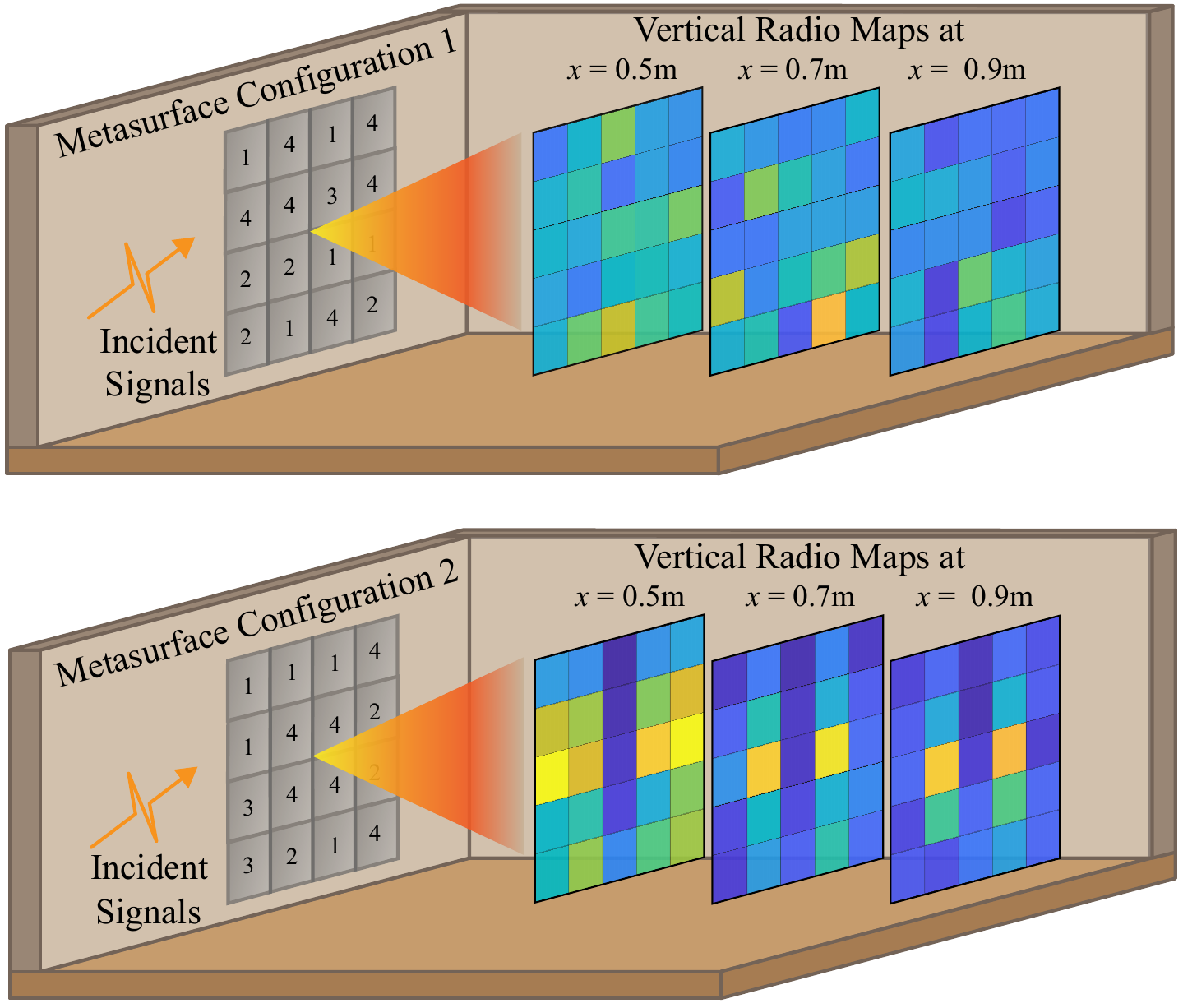}
  \caption{Constructing a radio map in 3D space through metasurface's capability of changing radio environments. The numbers on the metasurface denote the states of elements, and the radio maps are formed by the RSS values measured on the vertical plane in front of the metasurface, where a light (e.g., light yellow), or dark (e.g., dark blue) color indicates a high, or low RSS value. The size of each block is $10$cm. The
 radio maps are measured at three different planes/locations, i.e., $0.5$m, $0.7$m, and $0.9$m, in front of the metasurface. Note that The visualized radio maps measured at $x = 0.5$m can be easily distinguished from those radio maps measured at $x = 0.7$m or $0.9$m. }
  \label{f_received_signals}
\end{figure}

The radio environment reconfiguration capability of the metasurface can be described as follows.
Consider a metasurface with $M$ elements.
Given a incident signal on the metasurface with frequency $f_c$ and incident angle $\bm \varphi^I$, based on~\cite{tang2019wireless}, the reflected signals at a location can be calculated by
\begin{equation}
\label{equ: reflect signal}
y(\bm c) =  \sum_{m\in[1,M']} \frac{\lambda_c \exp(-j2\pi d_m/\lambda_c)}{4\pi d_m}\cdot r(\bm \varphi^I, \bm \varphi^R_m, c_m)\cdot x_m,
\end{equation}
where $\bm c = \{c_1, \cdots, c_M\}$ is the configuration of the metasurface, $c_m$ denotes the state of the $m$-th element. Here, $c_m \in \mathcal{C}_a,~\forall m \in \mathcal{M}$, where $\mathcal{C}_a = \{c^a_1, \cdots, c^a_{N_a}\}$ denotes the set of available states of an element and $N_a = 4$. $\lambda_c$ denotes the wavelength of the $f_c$ signals, $d_m$ is the distance from the $m$-th element to that location, and $x_m$ denotes the incident signal on the $m$-th element. Therefore, by changing the configuration of the metasurface $\bm{c}$, the reflected signals at different locations can be modified and the radio environment can be customized.

To illustrate how the metasurface configures the radio environment, we measure the $3$D radio map on three planes $0.5$m, $0.7$m, and $0.9$m in front of the metasurface given different configurations. 
The metasurface elements in MetaRadar have $4$ different states, and the incident signals are of frequency $f_c = 3.2$~GHz and incident angle $\bm \varphi^I = (60^\circ, 90^\circ)$.
Fig.~\ref{f_received_signals} shows the configurations and the corresponding measured RSS, where the number on the metasurface element denotes the corresponding state.
Besides, the color represents the value of the RSS at the center of each block, where a light (e.g., light yellow), or dark (e.g., dark blue) color indicates a high, or low RSS value.
The length of the block edge is $10$cm. 

Comparing the two visualized radio maps measured at $x = 0.5$m (or at another location), the RSS values at the same block are different, under the two different metasurface configurations. Meanwhile, the set of two radio maps measured at at $x = 0.5$m can be easily distinguished from another set of radio maps measured at $x = 0.7$m or $0.9$m. These indicate the ability of metasurface to customize the radio environment and further identify a certain location (e.g., $x = 0.7$m).

\subsection{RSS Modeling}
\label{ss_rsm}  

Since the reflectivity of the metasurface is determined by the configuration, the incident angles and the reflection angles, when the location of the AP is fixed, the RSS can be described by the user's location and the metasurface configuration. Suppose the user is located at one of the blocks in the SOI. The set of blocks in the SOI is denoted by $\mathcal{N} = \{1, \cdots, N\}$, and all the blocks have equal size with edge length $e$.

As shown in Fig.~\ref{f_architecture}~(a), the wireless channel between the AP and the user contains a line-of-sight (LOS) channel, $M$ reflected channels and multi-path channels. The $m$-th reflected channel is the reflection link through the $m$-th element. The multi-path channel accounts for the reflection and scattering in the indoor environment~\cite{Hashemi1993Indoor}.

Therefore, given emitted signal $x$, user's location $n$, and metasurface configuration $\bm{c}$, and assuming the superposition property of reflected channels~\cite{goldsmith2005wireless}, the signal received by the user can be expressed as
\begin{equation}
  y(\bm{c}, n) = h^{LOS}_n x + \sum_{m\in\mathcal{M}} h_{m, n}(c_m) x + h^{R}_n x + \xi,
  \label{calculation_y_n_c}
\end{equation}
where $h^{LOS}_n$ is the gain of the LOS channel, $h_{m, n}(c_m)$ is the channel gain of the $m$-th reflected channel, $h^{R}_n$ is the gain of the multi-path channel, and $\xi$ denotes the noise signal. Therefore, the RSS for configuration $\bm{c}$ and block $n$ can be expressed as $s(\bm{c}, n) = |y(\bm{c}, n)|^2$.

The distribution of the RSS can be approximated by the Gaussian distribution~\cite{youssef2005horus}, which can be expressed as
\begin{equation}
  \mathbb{P}(s(\bm{c}, n) = s) =  \dfrac{1}{\sqrt{2 \pi \sigma^2}} e^{-\dfrac{(s - \mu(\bm{c}, n))^2}{2\sigma^2}},
  \label{distribution_y_n_c}
\end{equation}
where $\sigma$ is the standard deviation, and $\mu(\bm{c}, n)$ is the mean of RSS for configuration $\bm{c}$ at block $n$. 

\subsection{Compressive Construction Technique}
\label{ssec: radio map construction}
According to (\ref{calculation_y_n_c}), we can calculate the mean RSS at any block for any configuration if the channel gains $h^{LOS}_n$, $\{h_{m, n}(c_m)\}$, and $h^{R}_n$ are known. However, it is difficult to measure the these channel gains directly. Instead, we measure the received signals for some critical configurations, which can be utilized to derive the RSS for any configuration.

Specifically, let $\bm{c}_{m, k}$ denote the critical configuration where the state of element $m$ is $c^a_k$ and the states of other elements are all $c^a_1$. After measuring the received signals under the configurations $\{\bm{c}_{m, k}\}, \forall m, k$, we can calculate the difference $\delta_{m, k}$, which is defined as
\begin{equation}
  \delta_{m, n, k} = h_{m, n}(c^a_k) - h_{m, n}(c^a_1) = y(\bm{c}_{m, k}, n) - y(\bm{c}_{m, 1}, n).
\end{equation}
Based on (\ref{calculation_y_n_c}), the received signal for $\bm{c}$ can be expressed as
\begin{equation}
  y(\bm{c}, n) = y(\bm{c}_{1, 1}, n) + \sum_{m \in \mathcal{M}} \delta_{m, n, k}.
\end{equation}
In total, there are $N_a \times M - M + 1$ critical configurations (exclude $M - 1$ repeated configurations). Using the received values under these critical configurations, we can derive radio maps for all the possible $N^M_a$ configurations.

However, in (\ref{calculation_y_n_c}) we assume the superposition property of the reflected channels, and it is not satisfied unless the mutual coupling among elements can be ignored. To reduce the mutual coupling, we group neighboring metamaterial units together, and design independent control and power supply circuits for each element, which is discussed in Section~\ref{s_i}. The performance of the proposed compressive technique is also evaluated in Section~\ref{ss_rrmc}.

\section{Fine-grained Localization Phase}
\label{s_ilp}

\begin{figure}[!t]
  \centering
  \includegraphics[width=3.3in]{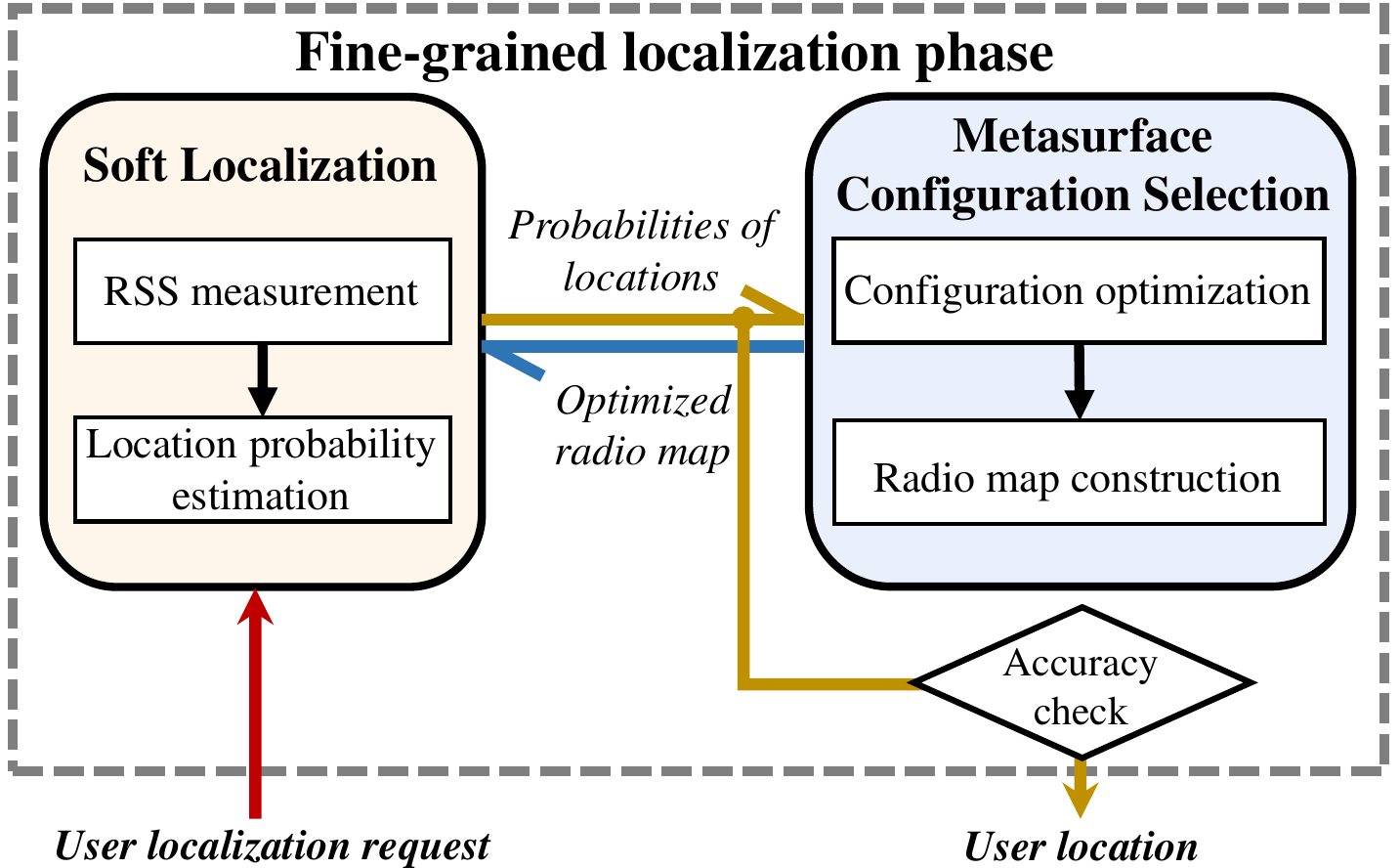}
  \caption{Illustration on the fine-grained localization phase.}
  \label{f_localizationPhase}
\end{figure}

In the fine-grained localization phase, MetaRadar locates multiple users with an acceptable localization accuracy by iteratively invoking two processes, i.e., \emph{soft localization} and {metasurface configuration selection}, as illustrated in Fig~\ref{f_localizationPhase}.

\subsection{Soft Localization}
The soft localization process is shown in the left part of Fig.~\ref{f_localizationPhase}.
As MetaRadar in the fine-grained localization phase receives a localization request from the user, it invokes the soft localization process first, which contains two tasks, i.e., the RSS measurement and location probability estimation.

\textbf{RSS measurement}: 
The RSS measurement takes place on the user device, which measures the RSS values of the received signals given the AP transmitting localization signals.
The user sends the measured RSS values along with its localization request to the AP.
Then, at the AP, it invokes the location probability estimation subprocess to calculate the probability of the user being at each block based on the user's measured RSS value and the current radio map.

\textbf{Location probability estimation}:
As the soft localization process is invoked iteratively, we denote the iteration index by $k$.
Then, the estimated probability for user~$i$ to be at the $n$-th block can be expressed as $p^k_{i, n}$, the optimized radio map in the $k$-th iteration can be expressed as $\bm \mu^{k}_{r}$, and the measured RSS value of user~$i$ can be expressed as $s^{k}_i$.
Based on the Bayes' theorem, the location probability estimation in the $(k+1)$-th iteration can be calculated as 
\begin{equation}
  p^{k+1}_{i, n} \approx \mathbb{P}(n | \bm \mu^{k}_{r}, s^{k}_i) = \dfrac{p^{k}_{i, n} \mathbb{P}(s^{k}_i | \bm \mu^{k}_{r}, n)}{\sum_{n\in\mathcal{N}} p^{k}_{i, n} \mathbb{P}(s^{k}_i | \bm \mu^{k}_{r}, n)},
  \label{d_prior_probability}
\end{equation}
As the initial condition, we assume that $p^0_{i, n}$ is equal for every block, i.e., $p^0_{i, n} = {1}/{N}, \forall n \in \mathcal{N}$, indicating that the location probabilities estimated in the $0$-th iteration are equal over all $N$ blocks.
After the location probabilities of users are obtained in each iteration, the AP decides whether the termination criterion is satisfied, which is discussed in Section~\ref{ss_tlp}.

\subsection{Metasurface Configuration Selection}
\label{ss_co}

\begin{algorithm}[!t]
  \caption{Configuration Optimization Algorithm}
  \label{a_coa}
  \KwIn{Estimated Probabilities $\{p_{i, n}\}$;}
  \KwOut{Configuration $\bm{c}$;}
  Initialize configuration $\bm{c}^1$ randomly, and calculate the corresponding RSS $\bm{\mu}^1$\;
  Set iteration index $z = 0$\;
  \While{$z \le Z_u$}{
    Set $k = k + 1$, $\bm{\mu}^{z+1} = \bm{\mu}^z$, and $\bm{c}^{z+1} = \bm{c}^z$\;
    Based on~(\ref{e_gradient}), generate the negative gradient $\bm{g}^z$ of the localization loss using $\bm{\mu}^z$\ and $\{p_{i, n}\}$\;
    \For {$m \in \mathcal{M}$}{
      Set $\bm{c}^* = \bm{c}^z$\;
      Enumerate all the possible values $c \in \mathcal{C}_a$ for element $m$ in $\bm{c}^*$\;
      Calculate the RSS $\bm{\mu}^*$ for $\bm{c}^*$ and the normalized difference $\bm{d}^*$ between $\bm{\mu}^*$ and $\bm{\mu}^z$\;
      If $|\bm{g}^z - \bm{d}^*| < |\bm{g}^z - \bm{d}^{z+1}|$ and the loss $l_u(\bm{c}^*) + \epsilon <  l_u(\bm{c}^z)$, then replace $\bm{c}^{z+1}$ with $\bm{c}^*$, and replace $\bm{\mu}^{z+1}$ with $\bm{\mu}^*$\;
    }
    If $\bm{c}^{z+1} = \bm{c}^z$, return $\bm{c}^z$\;
  }
\end{algorithm}

\textbf{Configuration optimization}: Based on the location probability of the users in current iteration, the metasurface configuration selection process optimizes the configuration of the metasurface to maximize the localization accuracy.
To evaluate the accuracy of localization, we define the \emph{localization loss} of the MetaRadar as follows.
Let $\mathcal{I} = \{1, \cdots, I\}$ denote the set of all the users who participate in the fine-grained localization, and the localization loss can be formulated as
\begin{equation}
  l(\bm{c}) = \sum_{i \in \mathcal{I}}\sum_{n \in \mathcal{N}} p_{i, n} \sum_{n' \in \mathcal{N}} \gamma_{n, n'} \int_{\mathcal{R}_{n'}} \mathbb{P}(s_i | \bm{c}, n) \cdot ds_i,
  \label{def_loss}
\end{equation}
where $p_{i, n}$ is the estimated probability that user $i$ is located at the $n$-th block. $\gamma_{n, n'}$ is the error parameter when the user is at the $n$-th block while the estimated location is the $n'$-th block, which is defined as
\begin{equation}
  \gamma_{n, n'} = |\bm{r}_n - \bm{r}_{n'}|,
\end{equation}
where $\bm{r}_n$ is the location of the $n$-th block center. $s_i$ is the RSS of user $i$, and $\mathcal{R}_{n'}$ is the decision region for block $n'$. That is, if $s_i \in \mathcal{R}_{n'}$, user $i$'s location is estimated as $n'$. Since most of decision regions $\mathcal{R}_{n'}$ are irregular, it is difficult to compute the localization loss in a closed form. In Appendix~\ref{a_ub}, an upper bound of localization loss $l_u (\bm{c})$ is provided. In the following, we use $l_u (\bm{c})$ in replace of $l (\bm{c})$ to reduce the computational complexity. Besides, we also eliminate the blocks with insignificant estimated probabilities, i.e., $\sum_{i \in \mathcal{I}} p_{i, n} \le \alpha$, to accelerate the computational speed.

Therefore, the configuration needs to be optimized to minimize the localization loss in this iteration. Generally, this problem is hard to solve due to the enormous number of available configurations and the complicated relationship between the configuration and the RSS. To solve this problem efficiently, we propose the configuration optimization algorithm. The basic idea is to use a modified gradient descent method which considers the limited states of metasurface elements to minimize the localization loss.

Specifically, the configuration $\bm{c}^1$ is initialized randomly at first, and we calculate the corresponding vector of RSS values $\bm{\mu}^1 = \{\mu(\bm{c}^1, 1), \cdots, \mu(\bm{c}^1, N^e)\}$. Here, the RSS vector only contains the values for blocks with estimated probabilities greater that $\alpha$, and the number of remaining blocks is $N^e$. In the $z$-th iteration, we first treat all the elements in $\bm{\mu}^z$ as continuous variables, and the minimization of localization loss can be viewed as an unconstrained minimization problem. Based on the idea of gradient descent method, we use the negative gradient $\bm{g}^z$ of the localization loss with respect to $\bm{\mu}$ as the search direction to the optimal result. According to Appendix~\ref{a_ub}, the $n$-th element in $\bm{g}^z$ can be expressed as
\begin{equation}
    \dfrac{\partial l_u}{\partial \mu_n} = \sum_{i, n'}(p_{i, n} + p_{i, n'}) \dfrac{\gamma_{n, n'}}{2}\left(-\dfrac{\mu_n - \mu_{n'}}{4\sigma^2}\right)e^{-\dfrac{(\mu_n - \mu_{n'})^2}{8\sigma^2}}.\label{e_gradient}
\end{equation}

Next, we adjust $\bm{\mu}^z$ in the direction of $\bm{g}^z$ by altering configuration $\bm{c}$. To be specific, we set $\bm{c}^* = \bm{c}^z$, and all the possible element states in $\bm{c}^*$ are enumerated successively. The corresponding RSS vector is denoted by $\bm{\mu}^*$, and the normalized difference $\bm{d}^* = |\bm{\mu}^* - \bm{\mu}^z|/|\bm{\mu}^*||\bm{\mu}^z|$. If $|\bm{g}^z - \bm{d}^*| < |\bm{g}^z - \bm{d}^{z+1}|$, which means the direction of $\bm{\mu}^* - \bm{\mu}^z$ is closer to the direction of $\bm{g}^z$ comparing to that of $\bm{\mu}^{z+1} - \bm{\mu}^z$, we will replace $\bm{c}^{z+1}$ and $\bm{\mu}^{z+1}$ with $\bm{c}^*$ and $\bm{\mu}^*$, respectively. Besides, we require $l_u(\bm{c}^*) <  l_u(\bm{c}^z)$ in order to assure the descent of localization loss. The iteration will end when no configuration element is changed or the iteration number $z$ exceeds $Z_u$.

The convergence of Algorithm~\ref{a_coa} can be analyzed as follows. Since the localization loss has lower bound $0$ and decreases at least $\epsilon$ with each iteration, the iteration cannot go on indefinitely, and the algorithm will converge. In addition, an upper bound $Z_u$ of iteration number is also set in order to terminate the algorithm within a limited period of time.

\textbf{Radio map construction}: After the optimal configuration is obtained by Algorithm~1, the AP constructs the corresponding radio map based on the method described in Section~\ref{ssec: radio map construction}, which is referred to as the optimized radio map $\bm \mu^{k}_{r}$. 
Different from the RSS vector, the optimized radio map contains the RSS values for all the blocks.
As shown in Fig.~\ref{f_localizationPhase}, the optimized radio map will be utilized in the location probability estimation process in the next iteration.

\subsection{Termination of the Localization Phase}
\label{ss_tlp}

Based on~(\ref{def_loss}), if $l_u(\bm{c}_k) < \beta_1 $ or $k > \beta_2$, the fine-grained localization phase will be terminated, where $\beta_1$ and $\beta_2$ denote the loss threshold and the maximal iteration number, respectively.
In this case, the index of the block where the location probability of the user is maximized is output as the user location.
The AP will estimate the locations of the users, and then send an ending signal to all the users together with their locations. 
Otherwise, the MetaRadar enters the next iteration and invokes the metasurface configuration selection process.

\section{Implementation}
\label{s_i}

In this section, we will provide detailed information on the metasurface, the Access Point (AP) and the user's module.

\subsection{Metasurface Module}

\begin{figure}[!t]
  \centering
  \includegraphics[width=3.3in]{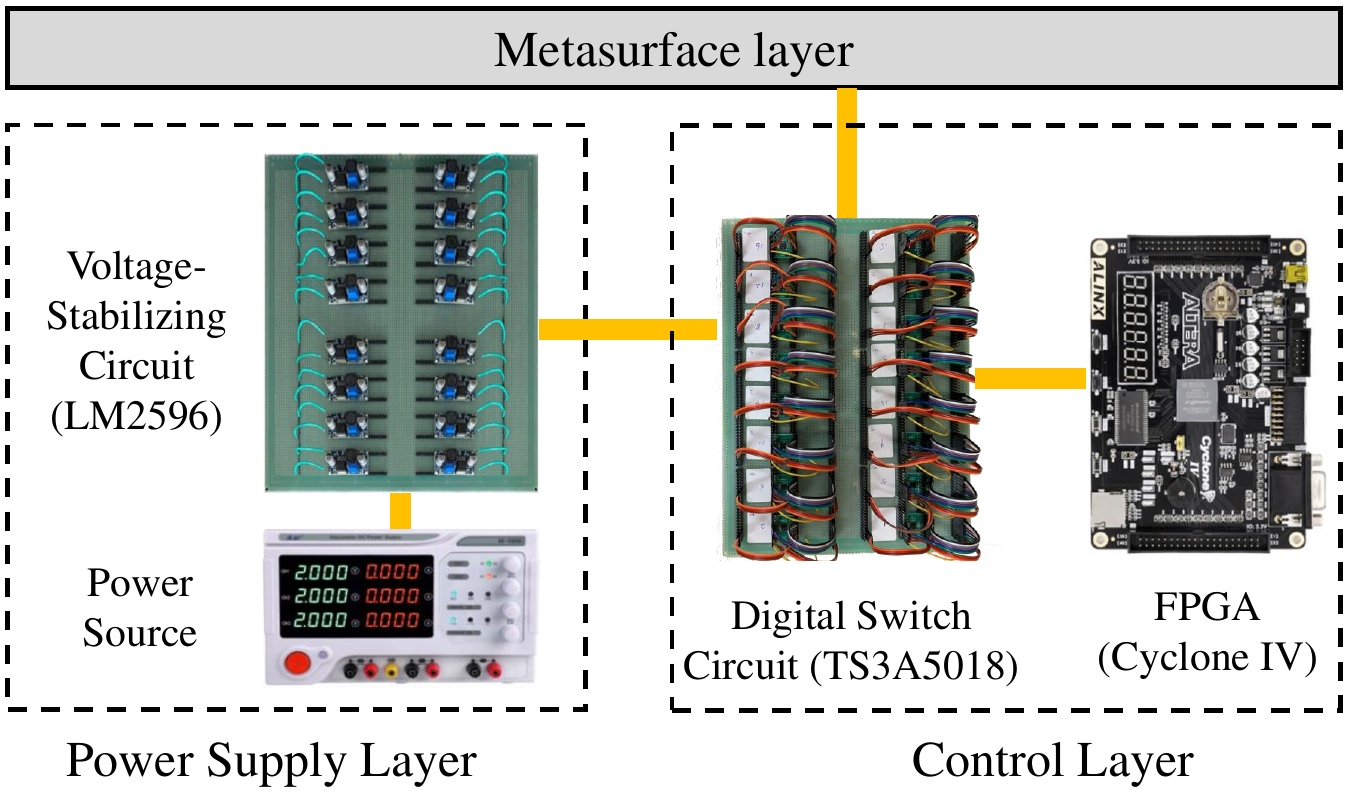}
  \caption{Three layers in the metasurface.}
  \label{f_metasurface_module}
\end{figure}

The structure of the metasurface module is illustrated in Fig.~\ref{f_metasurface_module}. It contains a metasurface layer, a control layer and a power supply layer. Specifically, the metasurface layer is utilized to reflect the RF signals. The control layer can modify the configuration of the metasurface to obtain the desired reflected waves. Finally, the power supply layer is used to provide stable electricity supply for the above layers.

\begin{figure}[!t]
  \centering
  \includegraphics[width=3.3in]{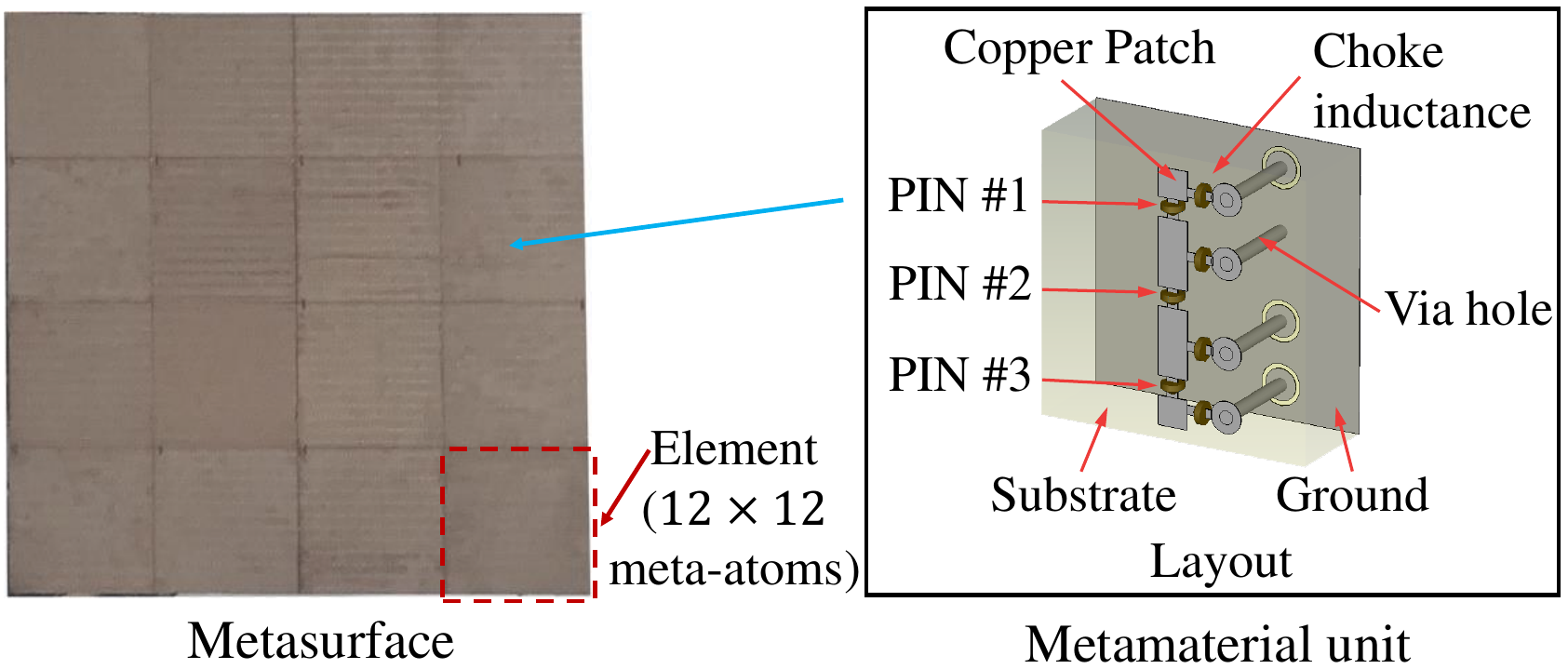}
  \caption{The formation of the metasurface with metamaterial units.}
  \label{f_metasurface_elements}
\end{figure}

As shown in Fig.~\ref{f_metasurface_elements}, the metasurface is a square sheet with the size $69 \times 69 \times 0.52$cm$^3$. It contains $16$ elements, and each element is composed of $12 \times 12$ metamaterial units. Each metamaterial unit has three sublayers~\cite{Li2019Machine}. On the top sublayer there are three copper patches which are connected by PIN diodes (BAR 65-02L). The substrate layer in the middle is made of Rogers 3010 with dielectric constant 10.2. A metallic layer is utilized as the ground at the bottom. The operation states of PIN diodes are controlled by the voltages applied on the via holes. When the applied bias voltage is $3.3$V(or $0$V), the corresponding PIN diode is at the ON (or OFF) state. The choke inductors ($30$nH) between the via hole and the via holes and the copper patches are utilized to separate the DC port and the RF signals.

\begin{table}[!t]
  \renewcommand{\arraystretch}{1.3}
  \caption{Reflectivity of a metamaterial unit in different states.}
  \label{t_r}
  \centering
  \begin{tabular}{|c|c|c|c|c|}
    \hline
    \textbf{Reflectivity} & \textbf{$c^a_1$} & $c^a_2$ & $c^a_3$ & $c^a_4$\\
    \hline
    \hline Amplitude & $0.95$ & $0.97$ & $0.93$ & $0.88$\\
    \hline Phase Shift & $-33^\circ$ & $60^\circ$ & $134^\circ$ & $-136^\circ$\\
    \hline
  \end{tabular}
\end{table}

The working frequency of the metasurface is at $3.2$GHz. Using CST Microwave Studio, Transient Simulation Package, the reflectivity of metamaterial unit in different states are calculated, and $4$ states are selected, which are denoted by $\{c^a_1, c^a_2, c^a_3, c^a_4\}$. As shown in Table~\ref{t_r}, these states have high amplitude of reflectivity, and the phase shifts between two adjacent states are close to $90^\circ$, which demonstrates the effectiveness of the metamaterial unit design.

To reduce the mutual coupling in the control layer and the power supply layer, we use a specific digital switch (TS3A5018) to control the bias voltages of all the metamaterial units in one element, and use the voltage-stabilizing circuit to supply the bias voltages. We use the FPGA (Cyclone IV) in the control layer to manipulate the digital switches and communicate with the AP.

\subsection{Access Point and User Modules}
The AP module consists of a laptop, a USRP, a low-noise amplifier and a horn antenna. The baseband signal generation, signal processing, configuration optimization and communication with FPGA are all performed on the laptop using python programs with GNU radio packet~\cite{blossom2004gnu}. The USRP (LW-N210) connected to the laptop can convert baseband signals to the RF signals with frequency $3.2$GHz and vise versa. Since we only utilize the signal strength of the received signals, the accurate synchronization among USRPs are not necessary. The RF signals generated by the USRP will first be amplified by the low-noise amplifier (ZX60-43-S+) and then sent to the directional double-ridged horn antenna (LB-800) which emits high gain RF signals to the metasurface~\cite{balanis2016antenna}.

The structure of the user module is similar to that of the AP module. The difference between these two modules is that we use a small polymer antenna FXUWB10 with the size $3.5 \times 2.45 \times 0.02$cm$^3$ in replace of the horn antenna in the user module in order to receive signals in all directions.

\subsection{Workflow Setting}
In the following we specify the workflow setting of the MetaRadar in practice.
The $3$D radio map preparation phase using the compressive construction technique is carried out first.
In this phase, the receiver antenna is placed at the center of each block in the SOI sequentially.
When the receiver antenna is in the $n$-th block, the metasurface changes to each of the $N_a\times M - M + 1$ critical configurations with an changing interval equaling to $0.5$s.
By this means, necessary measurement is obtained to derive the radio maps for all the possible $N_a^M$ configurations.

Then, MetaRadar enters the fine-grained localization phase and waits for the localization requests form users. 
The timeline is divided into cycles with duration $100$ms, and AP emits signals with frequency $f_c$ from the $30$ms to the $80$ms in every cycle.
The fine-grained localization will start if any user sends request to the AP in the last $20$ms in a cycle. The request contains the information of the average RSS value $\{s^0_i\}$ in this cycle. The time division multiplex (TDM) technique is adopted to separate the requests of different users. Specifically, the last $20$ms in every cycle is divided into $I_s$ time slots, where $I_s$ is the number of all the users, and each user sends its request signal during the assigned time slot.

After the localization phase starts, the MetaRadar will estimate the location probabilities, optimize the metasurface configuration, construct the radio map, and send the starting signals to the metasurface in the first $30$ms of the next cycle. The starting signal contains the optimized configuration $\bm{c}_1$.
After receiving the starting signals, the metasurface will change its configuration to $\bm{c}_1$. During the next $50$ms, the AP continuously emits sine wave signals with frequency $f_c$, and the users will record the value of the corresponding RSS and calculate the mean value. The mean value of RSS recorded by user $i$ is denoted by $s^1_i$. During the last $20$ms in this cycle, the users will send their values to the AP in a TDM manner. 

At the end of each iteration, the AP will decide whether the termination criterion is satisfied. 
Specifically, the loss threshold and the maximal iteration number are set to $\beta_1 = 0.1$ and $\beta_2 = 500$. The iteration will end if $l_u \le \beta_1$ or the iteration $k > \beta_2$, and users' location result will be output.

\section{Evaluation}
\label{s_e}

\begin{figure}[!t]
  \centering
  \includegraphics[width=3.3in]{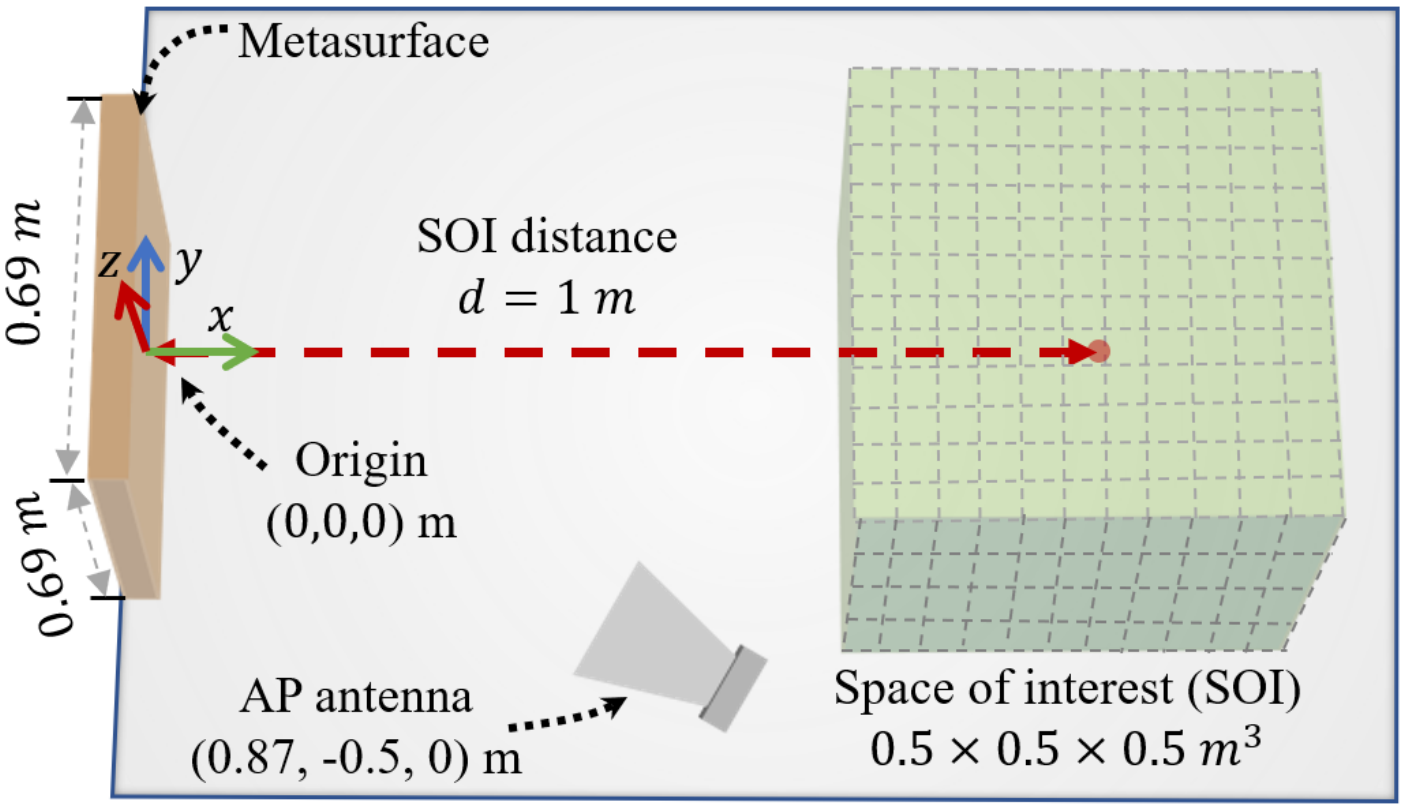}
  \caption{Experimental Layout.}
  \label{f_layout}
\end{figure}

In this section, we first introduce the setup of our test environment, and then present the evaluation results.

\subsection{Experimental Setup}

We perform the experiments in a classroom with size $25$m$^2$, and the walls of the classroom are made of bricks and concrete. As illustrated in Fig.~\ref{f_layout}, the metasurface is on the $y-z$ plane, and its center is located at $(0, 0, 0)$. Besides, the AP antenna is $1$m away from the metasurface and pointed at it. Moreover, the SOI is cuboid region with size $0.5 \times 0.5 \times 0.5$m$^3$. The center of the SOI is located at $(d, 0, 0)$, where $d$ denotes the distance between the SOI center to the metasurface. To evaluate the system performance for different distances, we choose three SOIs with distance $d = 1$m, $2$m, and $3$m, respectively. When building the radio map for each SOI, we discretize the SOI into blocks with size $5 \times 5 \times 5$cm$^3$ and record the signals for each block. There are no objects between the users and the metasurface.

To evaluate the performance of the localization system, the localization error is introduced. The localization error of each user is defined as the distance between the actual and the estimated location. For the whole system with multiple users, we use the average localization error of all the users as the localization error of the system. Here, we choose the center of the estimated block as the estimated location.

\subsection{Results for Radio Map Construction}
\label{ss_rrmc}

Since the residual mutual coupling among metasurface elements will influence the RSS model and the proposed method, in this subsection, we perform experiments to evaluate the performance of the proposed method. Specifically, we measure the amplitude and the phase of the received signals for the critical configurations and the RSS for $1,000$ random configurations. When measuring the received signals, the USRPs of the AP and the receiver USRSs are connected by the MIMO cable to realize the accurate timing between two USRPs, which is necessary to obtain the signal phase. We measure the received signals for $0.5$ seconds under each configuration, and the signal values are averaged over the whole period. The RSS for each configuration is measured for $50$ms. 

\begin{figure}[!t]
  \centering
  \includegraphics[width=3.1in]{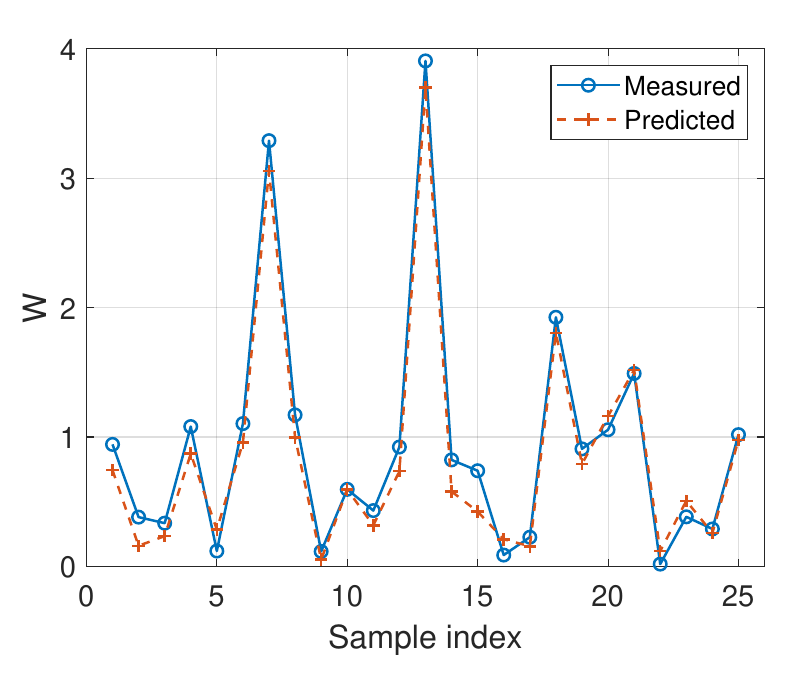}
  \caption{Measured RSS samples and the predicted RSS for different configurations.}
  \label{f_verify_d1}
\end{figure}

\begin{figure}[!t]
  \centering
  \includegraphics[width=3.1in]{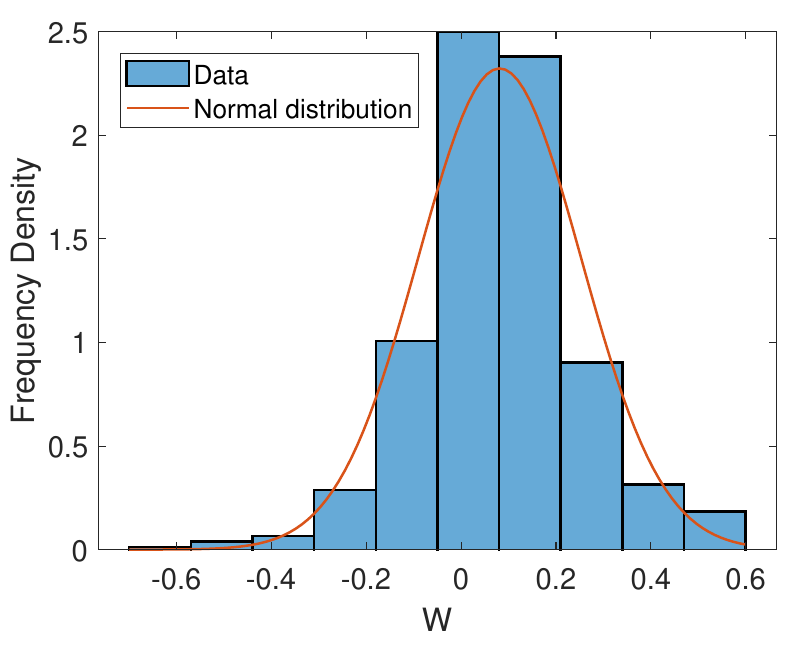}
  \caption{The distribution of deviation between measured RSS and predicted RSS and the normal distribution fit for the deviation.}
  \label{f_verify_d2}
\end{figure}

Fig.~\ref{f_verify_d1} shows the measured RSS samples and the predicted RSS for $25$ different configurations when the user is located at $(0.5, 0, 0)$m. We can observe that the measured RSS and the predicted RSS are very close. Fig.~\ref{f_verify_d2} presents the distrubution of deviation between the measured RSS and predicted RSS for $1,000$ different configurations. We can observe that the deviation approximately follows the normal distribution with mean $0.0811$W and standard deviation $0.1717$W, which are relatively small comparing to the average RSS $1.7753$W, indicating the effectiveness of the proposed compressive technique.

\subsection{Results for Single User Localization}

\begin{figure*}[!t]
  \centering
  \subfloat[]{
    \includegraphics[width=1.7in]{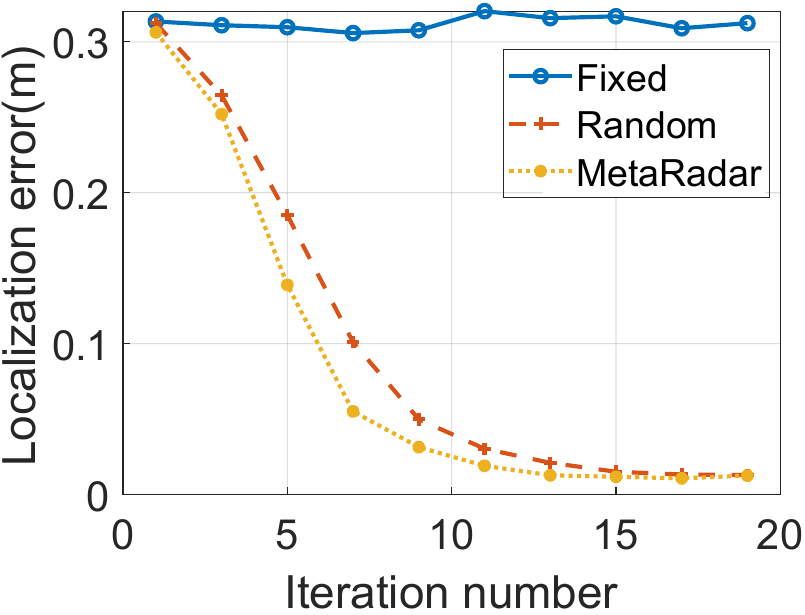}}
  \hspace{0.05in}
  \subfloat[]{
    \includegraphics[width=1.65in]{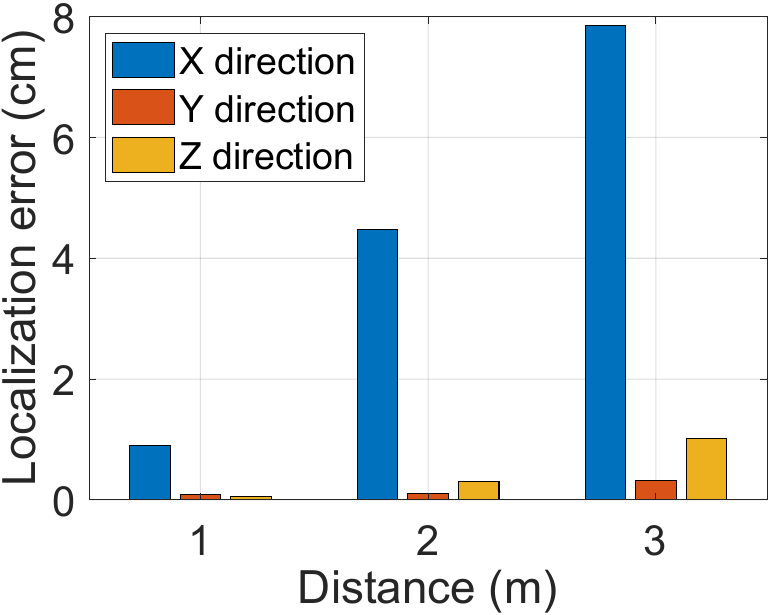}}
  \hspace{0.05in}
  \subfloat[]{
    \includegraphics[width=1.7in]{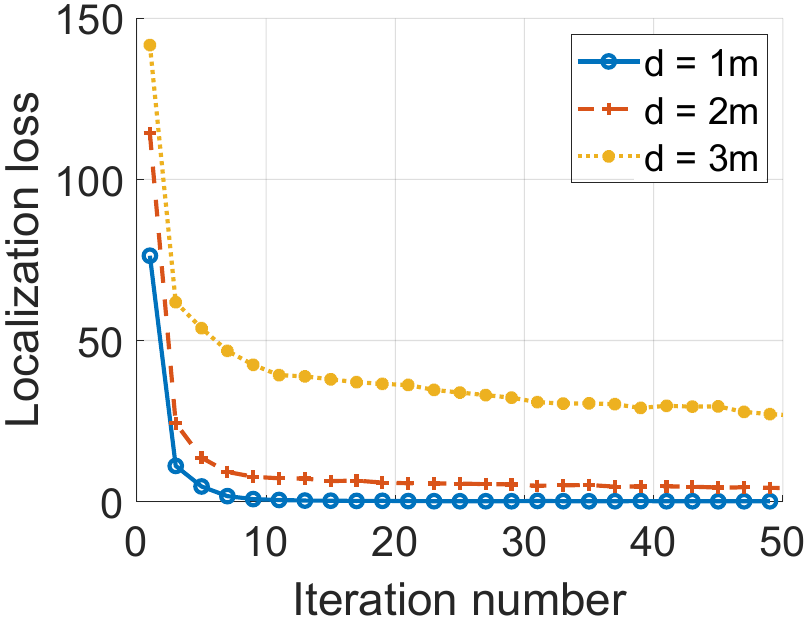}}
  \hspace{0.05in}
  \subfloat[]{
    \includegraphics[width=1.7in]{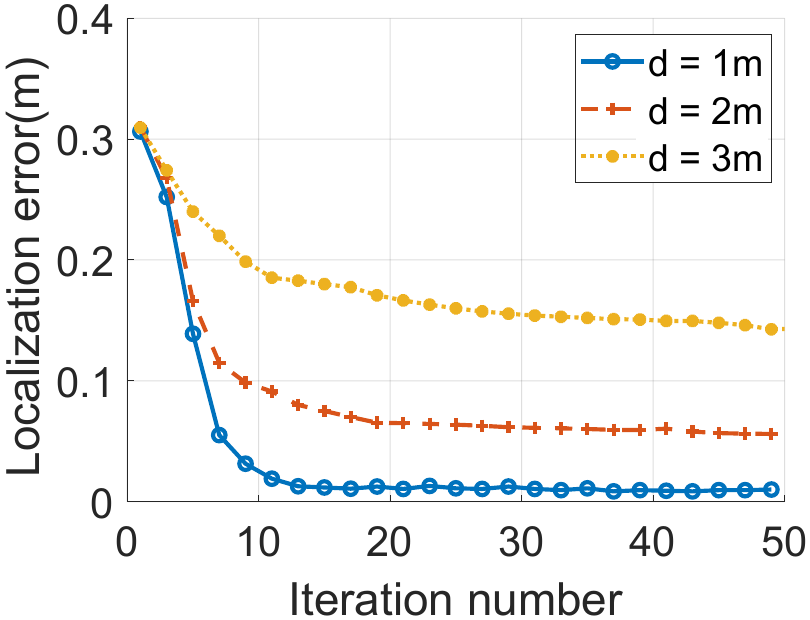}}
  \caption{The performance of fine-grained localization for single user: (a)~Three configuration algorithms; (b)~Localization errors in different directions versus distance; (c)~Localization loss versus iteration number; (d)~Localization error versus iteration number.}
  \label{f_single}
\end{figure*}

In this subsection, we present the experimental results for single user localization. 

Fig.~\ref{f_single} illustrates the performance of fine-grained localization for single user. To evaluate the performance of the MetaRadar, in Fig.~\ref{f_single}~(a), we also give the performance obtained by another two schemes when the distance from the SOI center to the metasurface $d = 1$m and the user is located at $(1, 0, 0)$. In the fixed configuration scheme, the states of all the metasurface elements are $c_1^a$. And in the random configuration scheme, random configurations are generated in different cycles. We can observe that the localization error $l_e$ of the fixed algorithm fluctuates between $0.30$m and $0.32$m, while the localization errors of the other two algorithm decrease when the number of cycles $n_c$ increases. We can also observe that the localization using the proposed algorithm has a faster decline speed compared to other two schemes, which has verified the effectiveness of the proposed algorithm.

Fig.~\ref{f_single}~(b) presents the localization error $l_e$ versus the distance from SOI center to the metasurface $d$ after $500$ iterations. We can observe that the localization error increases with the distance, and the proposed system can achieve a centimeter error when the distance between the metasurface and the user is smaller than $2$m. We can also observe that the localization error in the $x$ axis is clearly larger than those in the $y$ and $z$ axes. Since the $x$ axis is perpendicular to the metasurface, the correlation of signals in the $x$ direction is higher than those in the $y$ and $z$ directions, and therefore it is more difficult to distinguish different blocks in the $x$ direction.

\begin{figure*}[!t]
  \centering
  \subfloat[]{
    \includegraphics[width=2.35in]{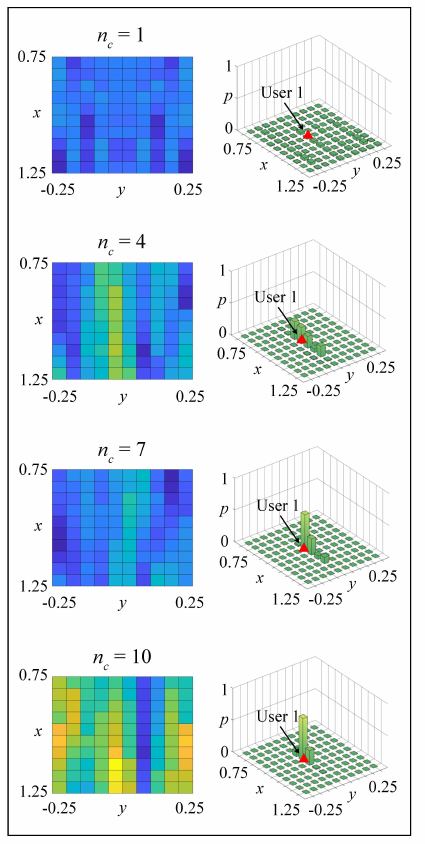}}
   \subfloat[]{
    \includegraphics[width=2.35in]{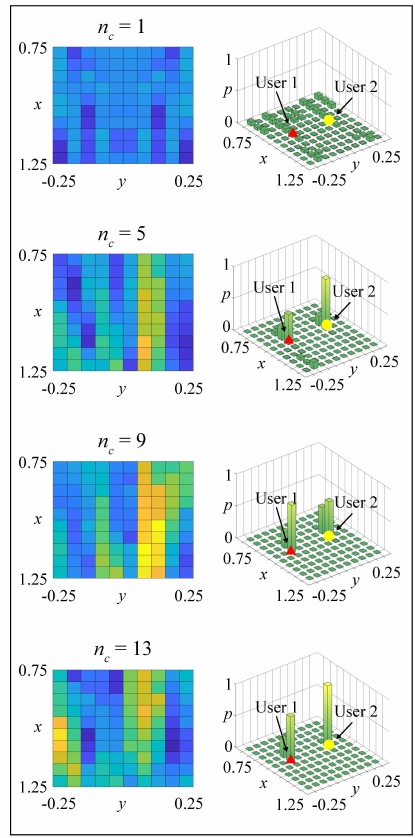}}
   \subfloat[]{
    \includegraphics[width=2.35in]{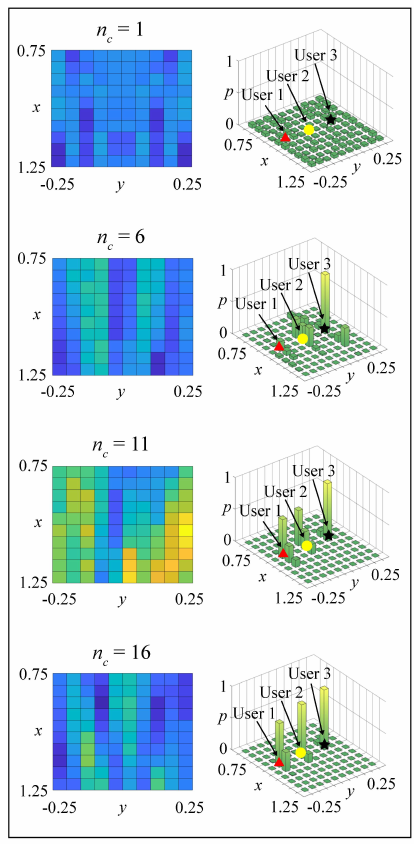}}
  
  \caption{Illustrations of the three-user localization process without obstruction: (a) One user; (b) Two users; (c) Three users. The first column in each subfigure shows the radio maps after different number of algorithm iterations, and the second column shows the corresponding probability distribution for different locations. Here, the probability is the sum of the probabilities of all the users. The ground truth of three users' locations are denoted by the red triangle, yellow circle, and the black star, respectively.}
  \label{f_example}
\end{figure*}

Fig.~\ref{f_single}~(c) depicts the localization loss $l_u$ versus the iteration number $n_c$. We can observe that the localization loss first decreases and then remains constant when the iteration number increases, which implies that the localization loss has a lower bound. We can also observe that for the same iteration number, the localization loss increases with the distance between the metasurface and the SOI center. Besides, in Fig.~\ref{f_single}~(d), we present the localization error $l_e$ versus the iteration number $n_c$. Similar to the results in Fig.~\ref{f_single}~(c), the localization error decreases when the iteration number increases, and increases with the distance. This implies that the localization error is positively correlated to the localization loss, which verifies the effectiveness of choosing localization loss to evaluate the accuracy of the localization algoirithm of MetaRadar.

\subsection{Results for Multiple User Localization without Obstruction}

\begin{figure*}[!t]
  \centering
  \subfloat[]{
    \includegraphics[width=1.65in]{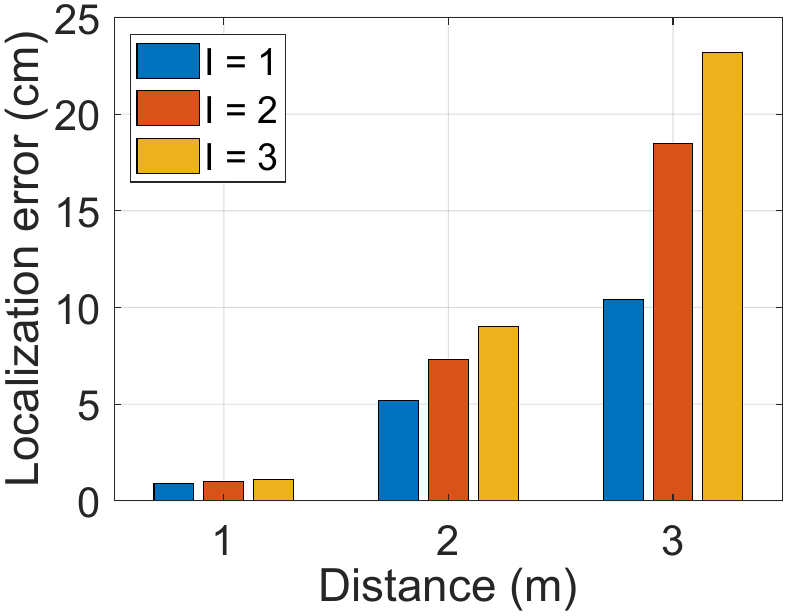}}
  \hspace{0.05in}
  \subfloat[]{
    \includegraphics[width=1.7in]{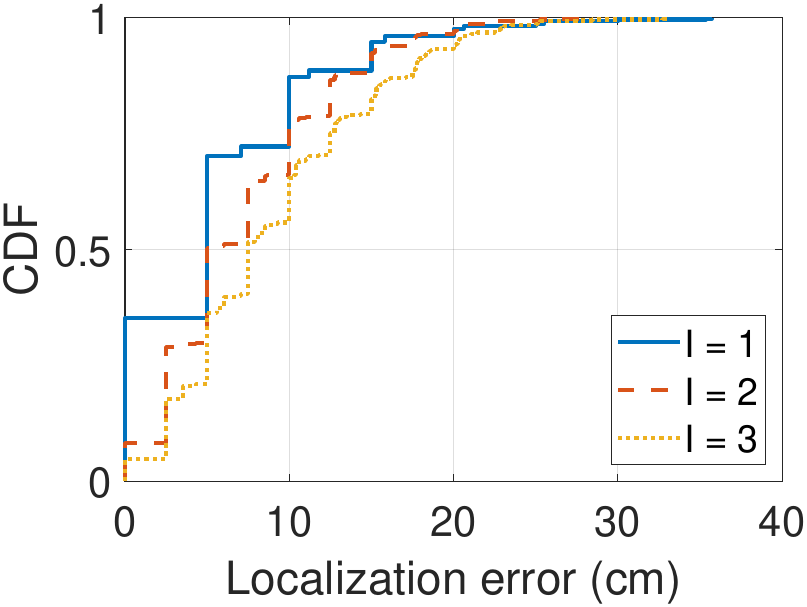}}
  \hspace{0.05in}
  \subfloat[]{
    \includegraphics[width=1.65in]{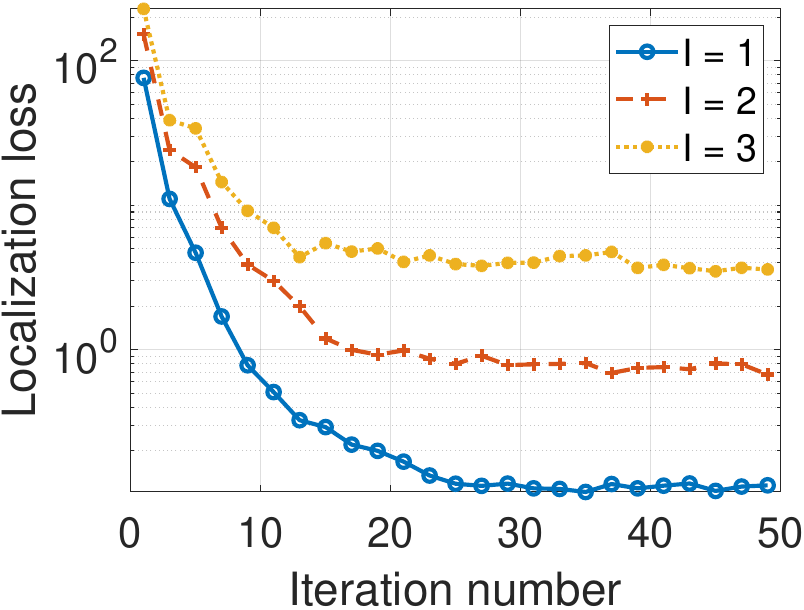}}
  \hspace{0.05in}
  \subfloat[]{
    \includegraphics[width=1.65in]{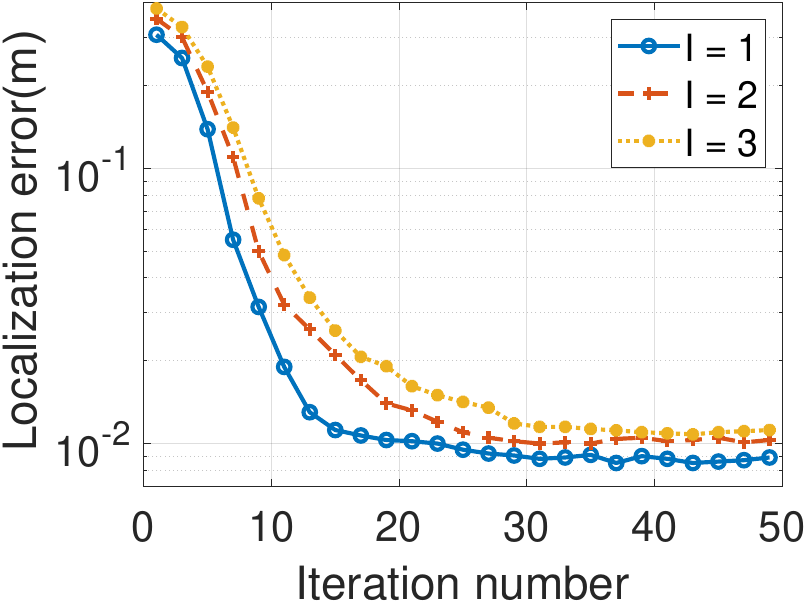}}
  \caption{When the number of users $I$ = 1, 2, and 3, the performance of fine-grained localization for multiple users without obstruction: (a)~Localization error versus distance; (b)~The CDF of localization error; (d)~Localization loss versus iteration number; (e)~Localization error versus iteration number.}
  \label{f_multi_no}
\end{figure*}

In this subsection, we evaluate the system performance for multiple user localization without obstruction. 

Fig.~\ref{f_example} illustrates the process of multi-user localization without obstruction. For display simplicity, we choose a planar area for localization. It is on the plane $z = 0$ with size $0.5 \times 0.5$m$^2$, and the distance between the area center to the metasurface center is $1$m. The ground truth of users' locations are labeled in the figures by the red triangle (the first user), yellow circle (the second user), and the black star (the third user), respectively. We can observe the RSS varies for different iterations. The probabilities are approximately uniformly distributed in each location in the first iteration (first row of subfigures in Fig.~\ref{f_example}), while after several iterations, the probabilities of locations near the ground truth are obviously higher than those of other locations, which implies the effectiveness of the proposed scheme. We can also observe that the locations in the $x$ direction near the ground truth have higher probabilities than the locations in the $y$ direction, indicating that it is more likely to misjudge the $x$ coordinate than the $y$ coordinate of the user's location (similar to Fig.~\ref{f_single}~(b), the $x$ axis is perpendicular to the
metasurface, and the correlation of signals in this direction is higher than those in the $y$ and $z$ directions.). Besides, we
can observe that for two user localization, the probability at location $(0.975, 0.125, 0)$ when $n_c = 9$ is smaller than that when $n_c = 5$. The decline in the probability is mainly due to the disturbance from the noise.

Fig.~\ref{f_multi_no} shows the fine-grained localization for multiple users without obstruction. In order to avoid the obstruction among users, all the users are all located at the plane parallel to the metasurface. Since the LOS channel between the metasurface and each user is not obstructed by other users, the interference caused by the existence of other users can be ignored.

Fig.~\ref{f_multi_no}~(a) presents the localization error of the system $l_e$ versus the distance $d$ between the users' locations and the metasurface when the number of users $I = 1$, $2$, and $3$  after $500$ iterations. We can observe that the localization error increases with the distance to the metasurface. Besides, the localization error also increases with the number of users. Fig.~\ref{f_multi_no}~(b) shows the CDF of the localization error for different number of users when the distance $d = 2$m. It can also be observed that the localization accuracy decreases when the number of users increases. This is because the metasurface configuration needs to be optimized for multiple users simultaneously, and the average signal variance introduced by metasurface for each user drops, which degrades the performance of the localization system.

Fig.~\ref{f_multi_no}~(c) depicts the localization loss $l_u$ versus the iteration number $n_c$ when $d = 1$m, and Fig.~\ref{f_multi_no}~(d) illustrates the localization error $l_e$ versus iteration number $n_c$ with the same distance between the metausrface and the users. Similar to the results in Fig.~\ref{f_single}~(c) and (d), the localization loss and the localization error for multi-user localization are also positively correlated. Besides, when the number of users increase, the localization loss and the error have slower decline speed, which implies a longer time for convergence.

\subsection{Results for Multiple User Localization with Obstruction}
\label{ss_rmulo}

\begin{figure}[!t]
  \centering
  \subfloat[]{
    \includegraphics[width=1.6in]{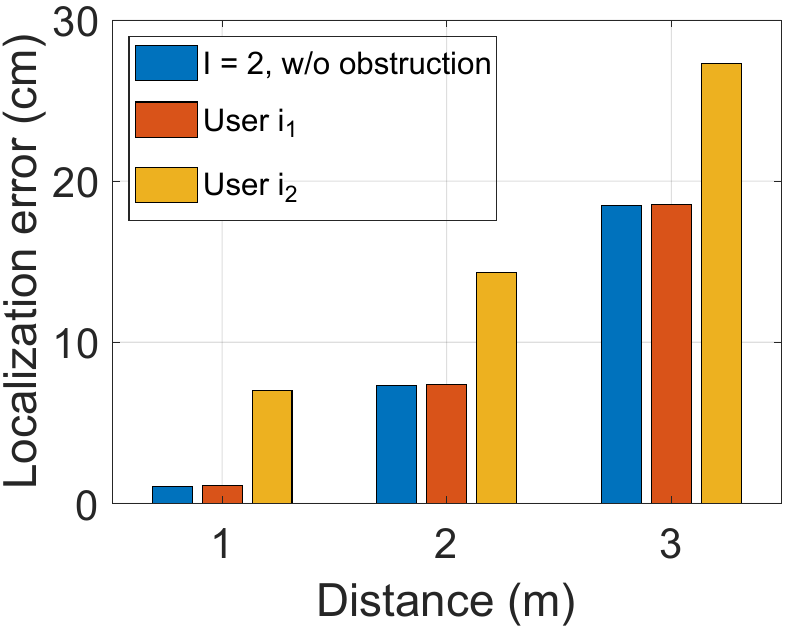}}
  \hspace{0.05in}
  \subfloat[]{
    \includegraphics[width=1.65in]{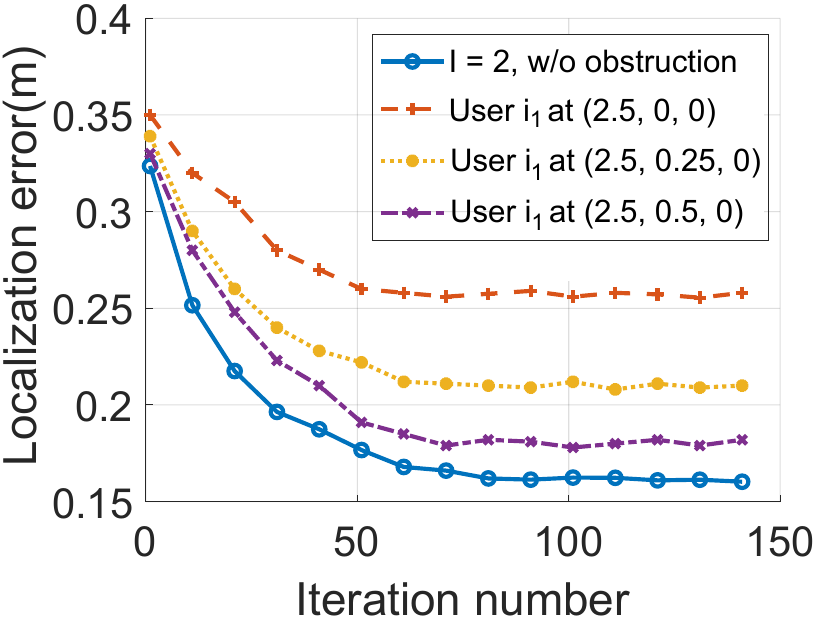}}
  \caption{The localization performance for two users with obstruction. The user closer to the metasurface is denoted by $i_1$, and the other user is denoted by $i_2$. "$I = 2$" in the legend is the localization error for two users without obstruction: (a)~Localization error versus distance; (b)~Localization error versus iteration number with different users' locations.}
  \label{f_obstruction}
\end{figure}

We present the experimental results in Fig.~\ref{f_obstruction} for multiple user localization with obstruction in this subsection. In Fig.~\ref{f_obstruction}~(a), we show the localization error $l_e$ versus the distance $d$ in this circumstance. Two users are placed in a line vertical to the metasurface, and the distance between the two users is $0.5$m. The user closer to the metasurface is denoted by $i_1$, and the other user is denoted by $i_2$. 
"$I = 2$" in the legend denotes the localization error without obstruction when the number of users is $2$. We can observe that at the same distance, the localization error of user $i_1$ is close to that without obstruction, and the localization error of user $i_2$ is visibly larger than others. This is because the existence of user $i_1$ will disturb the RF waves, and the received signals of user $i_2$ deviate from the results stored in the radio map.

Fig.~\ref{f_obstruction}~(b) presents the localization error $l_e$ of user $i_2$ versus the iteration number $n_c$. The location of user $i_2$ is fixed at $(3, 0, 0)$, while user $i_1$ is located at $(2.5, 0, 0)$, $(2.5, 0.25, 0)$, $(2.5, 0.5, 0)$, respectively. For comparison, we also give the result without obstruction when the number of users is $2$, which is denoted by $I = 2$ in the legend. We can observe that for the same iteration number, the localization error of user $i_2$ decreases when the distance in the $y$ direction between user $i_1$ and $i_2$ increases. This indicates that the localization error of a user is affected by the locations of other users.

\section{Discussion}
\label{s_d}

This paper investigates the indoor localization in a reconfigurable radio environment with metasurface. However, to achieve the ubiquitous localization assisted by metasurface, some additional challenges need to be addressed. 

\noindent \textbf{Interference among multiple users:} The fine-grained localization using the RSS based technique requires accurate radio maps. However, the existence of multiple users will induce new multi-path effect, and the corresponding received signals are different from the values stored in the radio map which is constructed by a single user. As shown in Section~\ref{ss_rmulo}, the localization accuracy will degrade if the LOS channel between the user and the metasurface is obstructed by other users. In future designs, some heuristic methods can be developed to adjust the radio map to improve the localization accuracy. Based on the estimated locations of multiple users, the effect of interference among multiple users can be incorporated.

\noindent \textbf{Support of a larger localization area:} According to the experimental results, the localization error increases with the distance between the metasurface and users. Therefore, the localization area is restricted comparing to conventional RSS based systems. There are two ways to increase the size of the localization area. The first way is to increase the size of the metasurface. The metasurface with larger reflection surface is able to control more reflected waves, and therefore a larger radio environment can be reconfigured for localization. The second way is to introduce more APs and metasurface to enhance the localization accuracy for different areas. Specifically, we can deploy APs and metasurface for every room in a building. Since the area of a room is comparable to the localization area of metasurface, in this way the localization accuracy of the whole building can be improved.

\noindent \textbf{Uncontrollable variance of radio environment:} Except the variance of metasurface reflection, the radio environment will also change with respect to time due to some uncontrollable factor, such as the movement of furniture, doors or any objects in the room. The system is unaware of the movements of objects, which is different from the interference among multiple users where the system knows the coarse locations of users. A research direction is to simultaneously locate the users and sense the variance of radio environment with the assist of metasurface. The sensing ability can be realized by imaging the indoor environment, which remains an active research topic of metasurface applications~\cite{hunt2014metamaterial}. Using the information of the sensed movements, the corresponding radio environment variance can be predicted to improve the localization accuracy.

\noindent \textbf{Narrow bandwidth of metasurface:} Since the reflectivity of the metasurface is sensitive to the frequency of the wireless signals, the reflectivity of a metamaterial unit for different states can be distinguished only for a very narrow bandwidth. Therefore, we use a single frequency for localization in this paper. The signals in narrow bandwidth contain less information of the radio environment than the wideband signals, where the latter can further improve the localization performance. However, the design of a wideband metasurface is still a challenging research topic. 

\section{Conclusion}
\label{s_c}

In this paper, we have proposed the metasurface assisted indoor localization system, MetaRadar. Unlike traditional RSS method, MetaRadar can control the RSS values in the radio environment by changing the configuration of the metasurface, which contributes to the improvement of localization accuracy. We have proposed the compressive construction method to build the radio map for all the possible radio environment, and designed the configuration optimization algorithm to select the favorable radio environment for fine-grained localization. MetaRadar has been implemented using USRPs and the experiments have been conducted in a classroom with different distances to the metasurface. The evaluation results have shown considerable improvement of the localization accuracy comparing to traditional RSS based systems with decimeter accuracy. Specifically, the proposed system equipped with a $0.48$m$^2$ metasurface can achieve a centimeter localization accuracy with up to $2$m localization range for single user and multiple users without obstruction. 

\appendices
\section{Derivation of the upper bound of the Localization Loss}
\label{a_ub}

We first utilize the union bound method to derive an upper bound for the integral in (\ref{def_loss}), which is quite tight for high signal-to-noise ratios~(SNRs)~\cite{proakis2007digital}. Specifically, we define the region $\mathcal{R}_{n', n}$ as
\begin{equation}
  \mathcal{R}_{n', n} = \{s_i : (s_i - \mu_{n'})^2 \le (s_i - \mu_{n})^2\}.\label{def_region}
\end{equation}
Since $\mathcal{R}_{n'} \subseteq \mathcal{R}_{n', n}$, we have
\begin{equation}
  \int_{\mathcal{R}_{n'}} \mathbb{P}(s_i | \bm{c}, n) \cdot ds_i \le \int_{\mathcal{R}_{n', n}} \mathbb{P}(s_i | \bm{c}, n) \cdot ds_i.
\end{equation}

Next, a closed form expression is provided for the upper bound. The region $\mathcal{R}_{n', n}$ can be expressed as
\begin{align}
  \mathcal{R}_{n', n} &= \{s_i : \mu^2_{n'} - 2s_i\mu_{n'} \le \mu^2_{n} - 2s_i \mu_{n} \}\notag\\
  &= \left\{s_i : s_i \mu^2_{n'} - s_i \mu^2_{n} - \mu_{n'} \mu_{n} + \mu^2_{n} \ge \dfrac{(\mu_{n'} - \mu_{n})^2}{2} \right\}\notag\\ 
  &= \{s_i : (s_i - \mu_{n})(\mu_{n'} - \mu_{n}) \ge 2 d^2_{n, n'}\},
\end{align}
where parameter $d_{n, n'}$ is
\begin{equation}
  d_{n, n'} = \dfrac{|\mu_{n'} - \mu_n|}{2}.
\end{equation}
Since $(s_i - \mu_{n})(\mu_{n'} - \mu_{n})$ follows Gaussian distribution with mean being $0$ and variance being $\sigma^2 (\mu_{n'} - \mu_{n})^2$, we have
\begin{align}
  \int_{\mathcal{R}_{n', n}} \mathbb{P}(s_i | \bm{c}, n) \cdot ds_i &= Q\left(\dfrac{2 d^2_{n, n'}}{\sigma |\mu_{n'} - \mu_{n}|}\right) \le \dfrac{1}{2} e^{-\dfrac{d^2_{n, n'}}{2\sigma^2}},
\end{align}
where the relationship $Q(x) \le \dfrac{1}{2} e^{-x^2/2}$ is utilized in the last step. Therefore, the upper bound of the localization error can be expressed as
\begin{align}
  l(\bm{C}) \le l_u(\bm{C}) = \sum_{i\in\mathcal{I}}\sum_{n\in\mathcal{N}} p_{i, n} \sum_{n'\in\mathcal{N}} \dfrac{\gamma_{n, n'}}{2} e^{-\dfrac{d^2_{n, n'}}{2\sigma^2}}.\label{simplified1}
\end{align}


%

\ifCLASSOPTIONcaptionsoff
  \newpage
\fi

\end{document}